\documentclass[onecolumn]{els-mrw} 

\usepackage{amsmath,amssymb,amsfonts,amsthm,makeidx,graphicx}
\usepackage{txfonts}
\usepackage{helvet}
\usepackage{tikz}
\usepackage{comment}

\usetikzlibrary{shapes.geometric, arrows.meta, positioning}

\usepackage{mathtools}
\usepackage{physics}          
\usepackage{bm}
\usepackage{bbm}
\usepackage{siunitx}
\usepackage{indentfirst}

\usepackage{booktabs, array, multirow}
\usepackage{tikz}
\usetikzlibrary{calc}
\usepackage{tikz-3dplot}
\usetikzlibrary{arrows.meta, calc, 3d}
\usepackage{xcolor}
\usepackage{ragged2e}
\usepackage[version=4]{mhchem}
\usepackage[colorlinks,linkcolor=blue,anchorcolor=blue,citecolor=blue]{hyperref}

\makeatletter
\@ifundefined{theorem}{\newtheorem{theorem}{Theorem}}{}
\@ifundefined{lemma}{}{}
\theoremstyle{definition}
\@ifundefined{axiom}{\newtheorem{axiom}[theorem]{Axiom}}{}
\@ifundefined{define}{}{}
\@ifundefined{assume}{}{}
\makeatother

\def\6{{\langle}}
\def\9{{\rangle}}
\def\Tr{\mathrm{Tr}}

\newcommand{\be}{\begin{equation}}
\newcommand{\ee}{\end{equation}}

\begin{document}
\Large 
\chapter{Generalised Probabilistic Theories}\label{chap:gpt}

\author[1]{Zujin Wen}
\author[1]{Zihan Wang}
\author[2,3]{Xiangjing Liu}
\author[1,4]{Oscar Dahlsten}

\address[1]{\orgname{Department of Physics, City University of Hong Kong},
  \orgdiv{Physics}, \orgaddress{Tat Chee Avenue, Kowloon, Hong Kong SAR, China}}
\address[2]{\orgname{Nanyang Quantum Hub}, \orgdiv{School of Physical and Mathematical Sciences, Nanyang Technological University}, \orgaddress{637371, Singapore}}
\address[3]{\orgname{Centre for Quantum Technologies}, \orgdiv{Nanyang Technological University}, \orgaddress{637371, Singapore}}
\address[4]{\orgname{Institute for Nanoscience and Applications}, \orgdiv{Southern University of Science and Technology}, \orgaddress{518055, Shenzhen, China}}


\articletag{}

\maketitle

\begin{abstract}
\large
We give an introduction to research associated with the generalised probabilistic theories framework, also known as the convex framework. States are real vectors representing lists of probabilities of measurement outcomes. Convex combinations of the vectors represent probabilistic combinations of different state preparations. Transformations are real matrices. Measurement outcomes are represented by functionals of the states, inner products of the state with a real vector, whose values are the probability of the measurement outcome in question. The framework generalises quantum theory. We describe the operational meaning of the framework, and how the concepts can be defined in terms of cones of states and measurement outcome vectors. We describe how the classical and quantum probability theories are represented in the framework. We describe Bell non-locality and the theory with super-quantum non-locality known as box world. We discuss generalised Hamiltonian mechanics in the discrete case and in continuous phase space, including the role of negativity of the phase space density in contextuality and tunnelling.     
\end{abstract}



\section{Introduction to GPT framework}
We now give a first introduction to a systematic framework for probabilistic theories within which quantum and classical are special cases. This is a sub-field of research in the quantum information community; for review papers see also~\cite{muller2021probabilistic,plavala2023general, janotta2014generalized, barnum2015post}. Understanding the framework  gives a deeper understanding of quantum theory, what its operational meaning is, how it relates to classical probability theory and what may lie beyond it. 

\subsection{Warm-up: Bloch sphere and beyond it}
As a soft warm-up, consider the Bloch sphere of 2-level quantum systems. With the Bloch sphere we represent the state as a {\em real} vector whose entries are {\em operational} quantities: 
\begin{equation}
\label{eq:blochvector}
\vec{\rho}_{Bloch}=\begin{pmatrix}
\langle X \rangle \\
\langle Y \rangle \\
\langle Z \rangle,
\end{pmatrix}
\end{equation}
the vector of expectation values of Pauli matrices $X$,$Y$ and $Z$. Other such representations are possible. For example, instead of $\langle g \rangle$ we could have listed $p(g=+1)$ and $p(g=-1)$, yielding an operational 6 dimensional real-vector state representation. That representation is one-to-one with the Bloch sphere state augmented to include the normalisation as a degree of freedom in that one can convert between the two with matrices: 
\begin{equation}
\begin{pmatrix}
p(X=+1) \\
p(X=-1) \\
p(Y=+1) \\
p(Y=-1) \\
p(Z=+1) \\
p(Z=-1)
\end{pmatrix}
=
\begin{pmatrix}
1/2 & 1/2 & 0 & 0 \\
1/2 & -1/2 & 0 & 0 \\
1/2 & 0 & 1/2 & 0 \\
1/2 & 0 & -1/2 & 0 \\
1/2 & 0 & 0 & 1/2 \\
1/2 & 0 & 0 & -1/2
\end{pmatrix}
\begin{pmatrix}
1 \\
\langle X \rangle \\
\langle Y \rangle \\
\langle Z \rangle
\end{pmatrix}.
\label{eq:blochtoprob}
\end{equation}
In the other direction, by inspection, a suitable $4\times 6$ matrix applied on the LHS of equation \ref{eq:blochtoprob} recovers the 4-dimensional Bloch vector.

Such real vector representations also allow us to write down states (together with measurements) which {\em violate quantum theory}, e.g. violate the uncertainty relation s.t. all the Paulis have expectation value 1:
\begin{equation}
\begin{pmatrix}
1 \\
0 \\
1 \\
0 \\
1 \\
0
\end{pmatrix}
=
\begin{pmatrix}
1/2 & 1/2 & 0 & 0 \\
1/2 & -1/2 & 0 & 0 \\
1/2 & 0 & 1/2 & 0 \\
1/2 & 0 & -1/2 & 0 \\
1/2 & 0 & 0 & 1/2 \\
1/2 & 0 & 0 & -1/2
\end{pmatrix}
\begin{pmatrix}
1 \\
1 \\
1 \\
1
\end{pmatrix}
\end{equation}
Those examples give some flavour of the framework of generalised probabilistic theories, and how it focusses on probabilities of outcomes.  

\subsection{Tables of data, states, measurements}
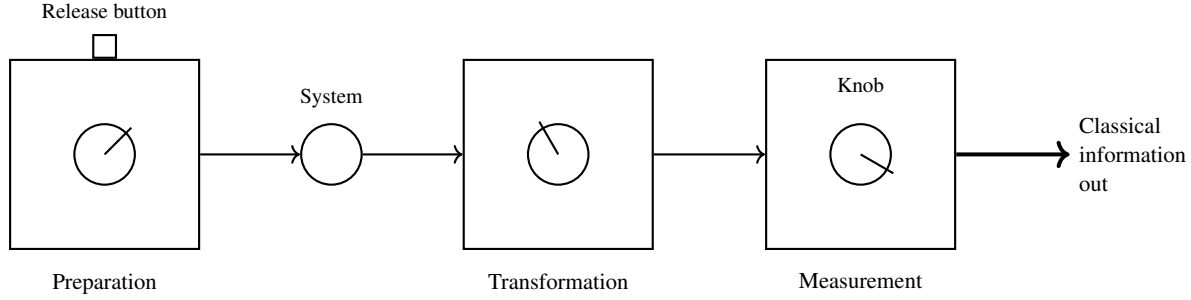
\begin{figure}\begin{tikzpicture}[
    box/.style={rectangle, draw, thick, minimum width=2.5cm, minimum height=2.5cm},
    circ/.style={circle, draw, thick, minimum size=0.8cm},
    arrow/.style={->, thick}
]
%
\node[box] (prep) at (0,0) {};
\node[circ] (prep-circle) at (prep.center) {};
\node[rectangle, draw, thick, minimum size=0.3cm, fill=white, anchor=south] (release) at (prep.north) {};
\node[above=0.1cm] at (release.north) {\small Release button};
\node[below=0.2cm] at (prep.south) {Preparation};

\node[circ] (system) at (3,0) {};
\draw[arrow] (prep.east) -- (system);
\node[above=0.1cm] at (system.north) {\small System};

\node[box] (trans) at (6,0) {};
\node[circ] (trans-circle) at (trans.center) {};
\draw[arrow] (system) -- (trans.west);
\node[below=0.2cm] at (trans.south) {Transformation};

\node[box] (meas) at (10,0) {};
\node[circ] (meas-circle) at (meas.center) {};
\draw[arrow] (trans.east) -- (meas.west);
\node[above=0.3cm] at (meas-circle.north) {\small Knob};
\node[below=0.2cm] at (meas.south) {Measurement};

\draw[arrow, line width=1.5pt] (meas.east) -- ++(1.5,0) node[right, align=left] {Classical\\information\\out};

\draw[thick] (prep-circle.center) -- ++(45:0.5);
\draw[thick] (trans-circle.center) -- ++(120:0.5);
\draw[thick] (meas-circle.center) -- ++(-30:0.5);

\end{tikzpicture}\caption{{\bf Hardy's depiction of the operational scenario.} The operational theory concerns probabilities of measurement outcomes. These probabilities depend on the choices, illustrated via knobs, of preparation, transformation and measurement.}
\label{fig:hardyoperational}
\end{figure}

Tables of data can be viewed as the essential basic representation of experimental records. The data refers to probabilities (relative frequencies) of outcomes with some labels. The process generating the data can be standardised~\cite{Hardy} as consisting of system preparation, transformation and measurement, followed by classical data output, as depicted in Figure~\ref{fig:hardyoperational}.

The data table generated by such an experiment is then well-defined. Let $M_i$ denote the i-th choice of measurement,  $o_j$ the j-th outcome, and $P_k$ the k-th preparation. Then running 
the experiment in Figure~\ref{fig:hardyoperational} many times, for different settings, generates a data table of the form of Table~\ref{tab:operational}.  
\begin{table}
\begin{center}
\begin{tabular}{ |l|c|c|c| } 
 \hline
${\bf M_i},{\bf o_j} $ $\backslash$ ${\bf P_k}$ & ${\bf P_1}$                    &       ${\bf P_2}$               & $\cdot \cdot \cdot$ \\ \hline
$M_1,o_1$                       & $p(o_1|M_1, P_1)$  & $p(o_1|M_1, P_2)$ & $  \cdot \cdot \cdot$ \\ \hline
$M_1,o_2$                       & $p(o_2|M_1, P_1)$  & $p(o_2|M_1, P_2)$ & $  \cdot \cdot \cdot$ \\ \hline
$\vdots$                          & $\vdots$                 & $\vdots$                  &  $\cdot \cdot \cdot$ \\ \hline
$M_2,o_1$                       & $p(o_1|M_2, P_1)$  & $p(o_1|M_2, P_2)$ & $  \cdot \cdot \cdot$ \\ \hline
$M_2,o_2$                       & $p(o_2|M_2, P_1)$  & $p(o_2|M_2, P_2)$ & $  \cdot \cdot \cdot$ \\ \hline
$\vdots$                          & $\vdots$                 & $\vdots$                  &  $\cdot \cdot \cdot$ \\ \hline
\end{tabular}
\caption{{\bf A general form for an experimental data table.} Probabilities are tabulated for given preparations $P$, measurements $M$ and outcomes $o$. Each column is the state for the given $P$.}
\label{tab:operational}
\end{center}

\end{table}

The initial {\bf state} of the system is associated with the preparation and can be represented by the column of the given preparation, which is a real vector as the entries are probabilities~\cite{mana2003can}. Sometimes we may choose for the table to be filled with expectation values 
for different measurements and preparations, in which case the entries are real numbers that may be negative. In either case the states are real vectors. We can call them $\vec{s}_k$ for example. 
The difference between expectation value vectors and probability vectors is cosmetic in that one can convert between them via a (pseudo) reversible matrix as in Eq.~\ref{eq:blochtoprob}. 

The {\bf measurement and outcome} pair $M_i, o_j$ can similarly be represented by the real vector that is all 0 except for a 1 at the $M_i, o_j$ row . We can call this vector $\vec{e}_{(i,j)}$, and then see 
\begin{equation}
\label{eq:effectstatedotprod}
p(o_j|M_i, P_k)=\vec{e}_{(i,j)}\cdot \vec{s}_k,
\end{equation}
where $\cdot$ is the dot product. Thus states and measurement outcomes can be represented as real vectors with the dot product between them giving us the probability of the outcome given the state and measurement choice. Eq.~\ref{eq:effectstatedotprod} is arguably the key rule in generalised probabilistic theories. 

$\vec{e}_{(i,j)}$ is often called the {\em effect} vector. This terminology goes back at least to Ludwig~\cite{ludwig1983foundations}. The effects correspond to (positive-operator valued) POVM elements in quantum theory. The effects can be viewed as more fundamental than the states in that in the data table one first lists effects before assigning probabilities to them via the state.

Effects can be treated as jointly forming a measurement only if the probabilities sum to one for all states. This is a generalisation of the so-called completeness condition for quantum measurements. Quantum measurement (POVM) elements sum to the identity. The generalisation is that measurement elements sum to the unit element $\vec{u}$, a vector that defines the normalisation. $\vec{u}\cdot \vec{s}=1$ for any normalised state $\vec{s}$. Thus the completeness condition, for effects to jointly form a complete measurement, reads 
\begin{equation}
\sum_k\vec{e}_{k}=\vec{u}.
\end{equation}

It is standard to assume that any probabilistic combination of allowed preparations is also allowed. If we prepare state $\vec{s}_k$ with probability $p_k$ one sees that, since $\sum_kp_kp(o_j|Mi,P_k)=\vec{e}_{(i,j)}\cdot \sum_kp_k\vec{s}_k$, the corresponding state vector is $\sum_k p_k\vec{s}_k$. 
(Probabilistic combinations are also called convex combinations and sometimes this approach to probabilistic theories is called the convex framework.) States which are not (non-trivial) probabilistic combinations of other states are called {\em pure} and the rest are {\em mixed}.

{\bf Transformations} alter the state and respect probabilistic mixtures. Altering the state means altering the probabilities and the transformations can therefore be treated as maps acting on the state vectors. Respecting probabilistic mixtures is to do with the physically sensible requirement that the transformation does not `know' the probabilities of given preparations and acts on the system in whatever way it was prepared: $T(p_1\vec{s}_1+p_2\vec{s}_2+...)= p_1T(\vec{s}_1)+p_2T(\vec{s}_2)+...$, where $T$ is the transform and state $\vec{s}_i$ was prepared with probability $p_i$,  consistent with $T$ being a real matrix given that the states are represented as real vectors.

\subsection{Framework in terms of state and effect cones}
Having discussed the operational meaning, we now give examples of common mathematical definitions of GPT concepts in terms of {\em cones} of real vectors, adopting part of the notation and definitions from Refs.~\cite{muller2021probabilistic,barnum2015post}. 

Let \( C \) be a subset of a vector space \( V \). The set \( C \) is called a \textbf{cone} if, for every vector \( x \) in \( C \) and every non-negative scalar \( \lambda \geq 0 \), the scaled vector \( \lambda x \) is also in \( C \):
\begin{equation}
    x \in C \text{ and } \lambda \geq 0 \implies \lambda x \in C
\end{equation}
A cone \( C \) is {\em convex} if it is also closed under convex combinations: 
\begin{equation}
    x, y \in C , \alpha, \beta \geq 0 \text{ and } \alpha + \beta = 1 \implies \alpha x + \beta y \in C
\end{equation}

\begin{definition}[State space]
A \emph{state space} is a pair $(A,\Omega_A)$ where $A$ is a real vector space of finite dimension and $\Omega_A \subset A$ is a compact, convex subset whose dimension is $ \dim \Omega_A = \dim A - 1$, such that there exists a linear functional, i.e., an element of the dual space $u_A\in A^*: A \rightarrow \mathbb{R}$, called the \emph{unit} (or \emph{normalisation}) functional, such that
      $  u_A(\omega) = 1$ for all $\omega \in \Omega_A .
    $
The associated cone of unnormalised states is defined by
$
    A_+ := \{ \alpha \omega \mid \alpha \ge 0,\ \omega \in \Omega_A \}.
$
Elements of $\Omega_A =\{\omega \in A_+| u_A(\omega) =1 \}$ are called \emph{normalised states}, while elements of $A_+$ represent states with arbitrary normalisation.
\end{definition}

The state space is convex as it is closed under convex combinations. A state $\omega \in \Omega_A$ is called \emph{pure} if it is an extremal point of the convex set $\Omega_A$, and otherwise \emph{mixed}. 

\begin{definition}[Measurements]
Let $(A, \Omega_A)$ be a state space. A linear functional, i.e., an element of the dual space $e\in A^*: A \rightarrow \mathbb{R}$, is an {\em effect} if it attains values between zero and one on all normalised states, i.e., $0\leq e(\omega) \leq 1$ for all $\omega \in \Omega_A$. An $n$-outcome measurement is a collection of effects $ e^{(1)},...,e^{(n)} $ with the property that $e^{(1)}+...+e^{(n)}=u_A$.
    
\end{definition}

The effect cone is $A^*_+= \{ \lambda  e| e \, \text{is an effect}, \lambda \geq 0 \}. $ Pure effects are extremal points of the convex set of effects.
Fiducial effects are a set of effects whose outcome statistics are sufficient to fully characterize any state.


\begin{definition}[Transformations]
Let $(A,\Omega_A)$ be a state space.  
A \emph{transformation} is a linear map
$
T : A \to A
$
such that
$
T(\Omega_A) \subseteq \Omega_A,
$
i.e.\ every normalised state is mapped to another normalised state.
A transformation $T$ is called \emph{reversible} if it is invertible and $T^{-1}$ is also a linear transformation.
A \emph{dynamical state space} is a triplet $(A,\Omega_A,\mathcal{T}_A)$, where $(A,\Omega_A)$ is a state space and $\mathcal{T}_A$ is a compact (or finite) group of reversible transformations.
\end{definition}

A transformation must map any valid input state
to another valid state, ensuring physical consistency of the output. For normalisation-preserving transforms then,
$
T(\Omega_A) \subseteq \Omega_A,
$
and
$
u_A \circ T = u_A.
$
Transformations may, like in quantum theory, lead to subnormalised states and the norm after the transform $u_A(T(\omega_A))$ is, like in quantum theory, the probability of the transform  taking place.  

\section{Classical probability theory}
\noindent Classical probability theory fits into the GPT framework. 

\subsection{Coin example} 
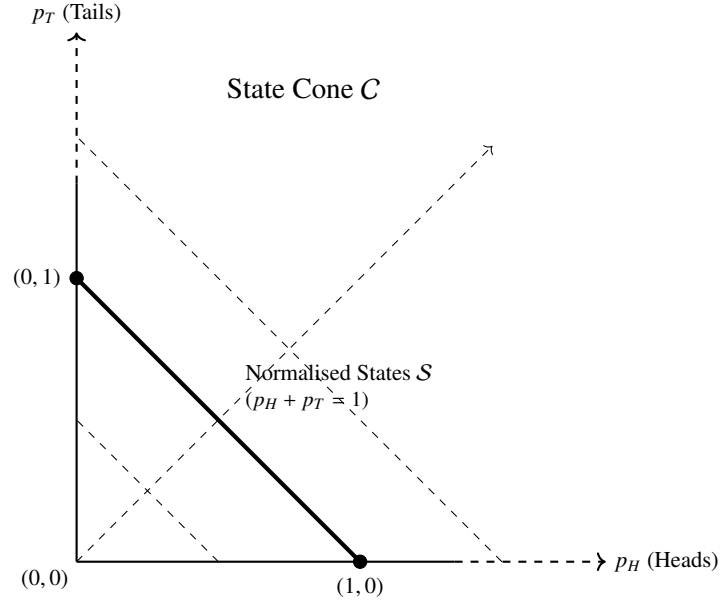
\begin{figure}[h]
    \centering
    \begin{tikzpicture}[scale=2.5]
        
        \draw[thick] (0,0) -- (2.0,0);
        \draw[thick, dashed, ->] (2.0,0) -- (2.8,0) node[right] {$p_H$ (Heads)};
        
        \draw[thick] (0,0) -- (0,2.0);
        \draw[thick, dashed, ->] (0,2.0) -- (0,2.8) node[above] {$p_T$ (Tails)};
        
        \draw[thin, dashed] (0.75,0) -- (0,0.75); 
        \draw[thin, dashed] (2.25,0) -- (0,2.25); 
        
        \draw[ultra thick] (1.5,0) -- (0,1.5);
        
        \filldraw[black] (1.5,0) circle (1pt) node[below=3pt] {$(1,0)$};
        \filldraw[black] (0,1.5) circle (1pt) node[left=3pt] {$(0,1)$};
        
        \draw[thin, dashed, ->] (0,0) -- (2.2,2.2);
        
        \node[font=\large] at (1.2, 2.5) {State Cone $\mathcal{C}$};
        
        \node[right] at (0.85, 1.0) {Normalised States $\mathcal{S}$};
        \node[right, font=\small] at (0.85, 0.85) {($p_H + p_T = 1$)};
        
        \node[below left] at (0,0) {$(0,0)$};
        
    \end{tikzpicture}
    \caption{{\bf The state cone and normalised states for a classical binary system.} The cone $\mathcal{C}$ is bounded by the $p_H$ and $p_T$ axes and extends infinitely outward, as indicated by the dashed lines. The normalised states $\mathcal{S}$ form a specific hyperplane slice (the thick solid line) that intersects the cone.}
    \label{fig:binary_cone}
\end{figure}

As an illustrative and generalisable example, consider a coin with Heads or Tails. Imagine three preparations, namely preparing heads ($P_H$), tails ($P_T$) or a uniform combination ($P_U$), and measuring just one measurement ($M_1$): Heads (H) vs Tails (T). The data table of the form of Table~\ref{tab:operational} is then
\begin{center}
\begin{tabular}{ |l|c|c|c| } 
 \hline
$M_1, o_j $ $\backslash$ $P_k$ & $P_H$                    &       $P_T$               & $P_u$ \\ \hline
$M_1,H$                                  & 1                            &            0                 & $  \frac{1}{2}$ \\ \hline
$M_1,T$                                  & 0                            &            1                 & $  \frac{1}{2}$ \\ \hline
\end{tabular}
\end{center}
Note that $P_u$ is equivalent to preparing $P_H$ and $P_T$ with equal probability: $\vec{s}_u=\frac{1}{2} \vec{s}_H+\frac{1}{2} \vec{s}_T,$ a mixed state. The states more generally have form $\vec{s}=\left( p(H) \,\,\, p(T) \right) ^T=p(H) \vec{s}_H+p(T)\vec{s}_T$. The unit effect is $\vec{u}=\left( 1 \,\,\, 1 \right) ^T$. The cone of states is depicted in Fig.\ref{fig:binary_cone}.

To see the dot product in action let the outcome representing Heads be  $\vec{h}=\left[1 \,\,\, 0\right]^T$ and similarly for tails $\vec{t}=\left[0 \,\,\, 1\right]^T$. Then 
\begin{eqnarray}
p(H)&=&\vec{h}\cdot \vec{s}\\
p(T)&=&\vec{t}\cdot \vec{s}
\end{eqnarray}

\subsection{General discrete case}
More generally we can define (discrete-case) classical probability theory as allowing for a measurement with some $d$ outcomes, where the pure states are those, like $\vec{s}_H$ and $\vec{s}_T$ above which have one definite outcome. Probabilistic combinations of the pure states are also allowed. For $d$ possible outcomes, the states have $d$ free parameters or $d-1$ if we take normalisation into account.
The pure states
$
\{\omega_1,\dots,\omega_d\} \subset \Omega_A,
$
can be written as $(1,0,0,...)^T,(0,1,0,...)^T...(0,0,0,...,1)$, with associated effects $\{e_1,\dots,e_d\} \subset A^*$ that can be represented via the same vectors as the pure states. The states $\{\omega_i\}_{i=1}^d$ are linearly independent; 
$\Omega_A$ is the convex hull (all the convex combinations) of $\{\omega_1,\dots,\omega_d\}$.

Transformations are commonly taken to be the stochastic matrices. Stochastic matrices are those whose entries are probabilities (real numbers between 0 and 1), and whose columns sum to 1 which can be shown to be necessary and sufficient for the matrix to take a normalised probability distribution to another normalised probability distribution. (One may also include matrices that correspond to conditioning on a given outcome and then the normalisation criterion is not required.)

\subsection{Only classical probability theory has unique decomposition of mixed states}
Classical probability theory is the only theory in the framework with unique decompositions of all mixed states. The pure states, of form $(1,0,0,...)^T,(0,1,0,...)^T...(0,0,0,...,1)$ are all linearly independent and form a complete but not overcomplete basis. Thus any state is a unique linear (and convex) combination of pure states. In contrast, when the set of pure states is an overcomplete basis there is linear dependence amongst the pure states. Linear dependence implies non-unique decomposition because if pure state 
\begin{equation}
\label{eq:purestatelinearlydependent}
\vec{\mu}_{i*}=\sum_ip_i\vec{\mu}_i+\sum_{j\neq i} n_j\vec{\mu}_j,
\end{equation}
 with $p_i>0$ and $n_i<0$ (some negative term must exist as $\vec{\mu}_{i*}$ is not a convex combination of other pure states) then we have two decompositions giving the same state:
\begin{equation}
\label{eq:nonunique}
\frac{1}{\sum_ip_i}\vec{\mu}_{i*}+\frac{1}{\sum_ip_i}\sum_{j\neq i}|n_j|\vec{\mu}_j=\frac{1}{\sum_ip_i}\sum_ip_i\vec{\mu}_i,
\end{equation}
which followed from rearranging of Eq.~\ref{eq:purestatelinearlydependent}  with a renormalisation factor to set the normalisation to 1\footnote{The normalisation of a state is the inner product with a `normalisation' vector $\vec{u}$. The LHS Eq.~\ref{eq:purestatelinearlydependent} and RHS have norm $\langle \vec{u}, \vec{\mu}_{i*}\rangle=1$. The RHS can be written as P+N where $P=\sum_ip_i>0$ and $N=\sum_{j\neq i}n_i<0$. Thus the LHS and RHS of Eq.~\ref{eq:nonunique} have the same normalisation, $1+|N|=P$, so dividing by P sets the normalisation to 1.}. The existence of at least one non-unique decomposition of a mixed state is often taken as the definition of a non-classical theory. Non-unique decomposition is connected to multiple `non-classical' properties of a theory including no-cloning and broadcasting~\cite{barnum2007generalized}.

If one imposes a restriction on the agent's knowledge in classical probability theory, a non-classical probability theory can be induced. Such a restriction--an epistemic restriction--can naturally be modelled by removing effect vectors, defining them to no longer be allowed. This can render certain preparations indistinguishable. For example consider two coins. In Table \ref{tab:full_joint}, the full joint measurements are allowed and there is a unique decomposition of the maximally mixed state. 
In Table \ref{tab:restricted_marginal} only measurements of individual coins are allowed and there is a non-unique decomposition of the maximally mixed state.
\begin{table}[htbp]
    \centering
    \begin{minipage}{0.48\textwidth}
        \centering
        \caption{Full joint measurements}
        \label{tab:full_joint} 
        \vspace{0.2cm}
        \resizebox{\textwidth}{!}{%
            \begin{tabular}{ |l|c|c|c|c|c| } 
            \hline
            Outcome \textbackslash{} Prep. & \(HH\) & \(HT\) & \(TH\) & \(TT\) & 
            \begin{tabular}{@{}c@{}}
                \(\frac{1}{4}HH + \frac{1}{4}HT\) \\ 
                \(+ \frac{1}{4}TH + \frac{1}{4}TT\)
            \end{tabular} \\ \hline
            \(H_1 H_2\)                    & 1    & 0    & 0    & 0    & \(\frac{1}{4}\) \\ \hline
            \(H_1 T_2\)                    & 0    & 1    & 0    & 0    & \(\frac{1}{4}\) \\ \hline
            \(T_1 H_2\)                    & 0    & 0    & 1    & 0    & \(\frac{1}{4}\) \\ \hline
            \(T_1 T_2\)                    & 0    & 0    & 0    & 1    & \(\frac{1}{4}\) \\ \hline
            \end{tabular}%
        }
    \end{minipage}\hfill
    \begin{minipage}{0.48\textwidth}
        \centering
        \caption{Restricted marginal measurements}
        \label{tab:restricted_marginal} 
        \vspace{0.2cm}
        \resizebox{\textwidth}{!}{%
            \begin{tabular}{ |l|c|c|c|c|c| } 
            \hline
            Outcome \textbackslash{} Prep. & \(HH\) & \(HT\) & \(TH\) & \(TT\) & 
            \begin{tabular}{@{}c@{}}
                \(\frac{1}{2}HH + \frac{1}{2}TT\) \\ 
                \(= \frac{1}{2}HT + \frac{1}{2}TH\)
            \end{tabular} \\ \hline
            \(H_1\)                        & 1    & 1    & 0    & 0    & \(\frac{1}{2}\) \\ \hline
            \(T_1\)                        & 0    & 0    & 1    & 1    & \(\frac{1}{2}\) \\ \hline
            \(H_2\)                        & 1    & 0    & 1    & 0    & \(\frac{1}{2}\) \\ \hline
            \(T_2\)                        & 0    & 1    & 0    & 1    & \(\frac{1}{2}\) \\ \hline
            \end{tabular}%
        }
    \end{minipage}
\end{table}

\section{Quantum theory in generalised probabilistic theory representation}
\label{subsec: quantum pauli}
We firstly describe the states and effects in quantum theory, then transformations.
\subsection{Data table, states and effects, for quantum theory}
\label{sec:DataTable}
In the standard quantum representation the state is the density matrix, and the measurement outcome (which shall later be generalised to POVM elements) represented by 
$\ket{a(i,j)}\bra{a(i,j)}$. 
Thus we can use the Born rule to populate the data table, according to
\begin{equation}
\label{BornRuleasGPT}
p(o_j|M_i, P_k)=\Tr (\ket{a(i,j)}\bra{a(i,j)}\rho_k).
\end{equation}

Finite-dimensional Quantum theory has a continuum of states and measurements  which makes it subtle whether a data table with columns representing states is sensible to write down. However the probability distributions consistent with quantum theory, whilst depending on continuous parameters, can be written as linear combinations of finite sets of basis functions.

A $d \times d$ quantum density matrix, or other hermitian operator, can be written as a linear combination of $d^2$ pure states. Consider \cite{Hardy} the following states:
\begin{equation}
    \left\{ \ket{x}, \overbrace{\frac{\ket{x} + \ket{y}}{\sqrt{2}}}^{\ket{+_{xy}}}, \overbrace{\frac{\ket{x} + i\ket{y}}{\sqrt{2}}}^{\ket{(+i)_{xy}}} \right\}_{x,y \in \{1, 2, \dots, d\}}
\end{equation}
The density matrix $\rho$ can be written as:
\begin{equation}
    \rho = \sum_{x,y} \rho_{xy} \ketbra{x}{y} \quad ; \quad \rho = \rho^\dagger \iff \rho_{ij} = \rho_{ji}^*
\end{equation}
Consider the following linear expansion in terms of pure states:
\begin{equation}
    \rho = \sum_{x} D_{xx} \ketbra{x}{x} + \sum_{y > x} \left( A_{xy} \ketbra{+_{xy}}{+_{xy}} + B_{xy} \ketbra{(+i)_{xy}}{(+i)_{xy}}\right)
\end{equation}
Expanding the terms in the sum over $y > x$, we get:
\begin{equation}
    \frac{A_{xy}}{2} \Big( \ketbra{x}{x} + \ketbra{x}{y} + \ketbra{y}{x} + \ketbra{y}{y} \Big)
\end{equation}
and 
\begin{equation}
    \frac{B_{xy}}{2} \Big( \ketbra{x}{x} + i\ketbra{y}{x} - i\ketbra{x}{y} + \ketbra{y}{y} \Big).
\end{equation}
Thus
\begin{align}
    \rho_{xx} &= D_{xx} + \sum_{y > x} \frac{A_{xy} + B_{xy}}{2}+\sum_{y < x} \frac{A_{yx} + B_{yx}}{2} \\
    \text{and} \quad \rho_{xy} &= \frac{A_{xy} - i B_{xy}}{2} \quad \text{for } y > x \\
    \text{and} \quad \rho_{yx} &= \frac{A_{xy} + i B_{xy}}{2} \quad \text{for } y > x.
\end{align}
The number of pure states in the expansion is $d^2$. There are $d$ of type $\ket{x}$. The number of times $y>x$ happens can be calculated as an arithmetic series sum $S=(d-1)+(d-2)+...1=(d-1)d/2$, and there are two states for each such case so $(d-1)d$ states from the $y>x$ part of the expansion. In total that is $(d-1)d+d^2=d^2$ states. There are also $d^2$ free real parameters in a Hermitian non-normalised density matrix ($d$ from diagonal, $(d-1)d$ from off-diagonals) so the expansion can be said to be optimal in that sense.  

The effects can be expanded in the same manner. The effects are the POVM (positive operator valued measure) elements $\Pi_{k}$, where $k$ labels the outcome. POVM elements generalise projectors $\ket{k}\bra{k}$. A POVM is a set of positive semidefinite (and thus hermitian) operators $\{M_i \}$ that satisfy the completeness relation $\sum_i M_i = \mathbb{I}$. A single POVM element $M$ can be expanded in the operator basis:
\begin{align}
M = \sum_j k_m \pi_j \equiv \vec{r}_M \cdot \vec{\Pi},
\end{align} 
Plugging linear expansions in terms of pure states (or other basis elements for hermitian operators), $\Pi_{k}=\sum_m k_m \pi_m$ and $\rho=\sum_l s_l h_l$, into the Born rule gives
\begin{equation}
\label{eq:BornRule}
p(k)=\Tr(\rho \Pi_k)=\sum_{m,l}k_ms_l\Tr(h_lg\pi_m):=\vec{k}^T D\vec{s},
\end{equation}
where $D_{lm}=\Tr(h_l\pi_m)$. $D$ is sometimes called the Gram matrix.

If $D_{lm}=c\delta_{lm}$ for some constant $c$, the Born rule corresponds up to a factor to the standard dot product. That is the case if we let the basis that $\pi_m$ and $h_l$ are part of be the $n$-qubit Paulis $\{X,Y,X,I\}^{\otimes n}$. For  $n$-qubit Paulis, $D_{lm}=\Tr(h_l\pi_m)=d\delta_{lm}$, where
$d=2^n$. Thus, in that case, one can visualise the Born rule as a dot product between real vectors, one representing the state and one representing the 
POVM element.



As an example of the dot product case associated with the Pauli basis, if one wants to calculate the probability of getting the outcome represented by vector $\ket{0}$ given a pure state $\ket{\psi}=c_0\ket{0}+c_1\ket{1}$ in this representation one would represent the matrix $\ket{0}\bra{0}$ with the vector
 $r_0=1/2$, $r_1=r_2=0$, $r_3=1/2$ 
 and the matrix $\ket{\psi}\bra{\psi}$ with the vector  
 $s_0=1/2$, $s_1=\Tr(\ket{\psi}\bra{\psi}X)/2$, $s_2=\Tr(\ket{\psi}\bra{\psi}Y)/2$, $s_3=\Tr(\ket{\psi}\bra{\psi}Z)/2$ such that $p(0)_{\ket{\psi}}=2\sum_i r_i s_i$.

\subsection{Transformations}
Transformations in quantum theory are described by completely positive and (when normalisation-preserving) trace-preserving (CPTP) maps $\mathcal{N}$ which change the probabilities of measurement outcomes. These become real matrices in the GPT framework.

To determine the real matrix in a systematic way, it is convenient to firstly note that an equivalent expression for POVM element $M$ as a real vector is
\begin{align}
\vec{p}_M := \Tr \big[ M\, \vec{\Pi}\big]
:= \begin{pmatrix}   
\Tr[M \pi_1] \\  
\Tr[M\pi_2]  \\
\vdots  \\
\Tr[M \pi_{d^2}]    
\end{pmatrix}
= D^T \vec{r}_M 
= D\,\vec{r}_M,
\end{align}
where again we used the Gram matrix $D$.

Accordingly,
\begin{align}
p_{\mathrm{meas}} 
= \sum_{i,j} b_j \, \Tr[\pi_j \pi_i] \, a_i 
= \vec{r}_M^{\,T} D \vec{r}_s,\\
= \vec{p}_M^{\,T} D^{-1} \vec{p}_s.
\end{align}

Let $\rho' = \mathcal{N}(\rho)$. Then, in the vector representation,
\begin{align}
\vec{p}'_s 
= \Tr\big( \vec{\Pi}\, \rho' \big)
= \Tr\big( \vec{\Pi}\,\mathcal{N}(\vec{\Pi}^T\vec{r}_s) \big) 
=\Tr\big( \vec{\Pi}\,\mathcal{N}(\vec{\Pi}^T) \big)\vec{r}_s
= \Tr\big( \vec{\Pi}\,\mathcal{N}(\vec{\Pi}^T) \big) D^{-1} \vec{p}_s 
:= Z\,\vec{p}_s,
\end{align}
where 
\begin{align}
Z := \Tr\big( \vec{\Pi}\,\mathcal{N}(\vec{\Pi}^T) \big) D^{-1}
\end{align}
is the real matrix representing the transformation $\mathcal{N}$ on the state vectors.

A special subset of transformations is the set of \textbf{reversible transformations}, denoted $\Gamma^{\text{reversible}}$, consisting of all invertible matrices $Z$ whose inverses also belong to $\Gamma$. These matrices form a representation of a group. Since this group is continuous and the vectors $\vec{p}$ generated by its action remain bounded, $\Gamma^{\text{reversible}}$ represents a compact Lie group. Every real representation of a compact Lie group is equivalent, under an appropriate change of basis, to a real orthogonal representation.  Under such a change, let $Z\in\Gamma$ be transformed to $Y\in\Omega$, and $\vec{p}_s$ to $\vec{q}_s$. The formula $p_{\text{meas}} = \vec{r}_M \cdot \vec{p}_s$ then becomes
\begin{align}
p_{\text{meas}} = \vec{r}'_M \cdot \vec{q}_s,
\end{align}
where $\vec{r}'_M$ is the transformed measurement vector. For reversible transformation $Y$, then  
\begin{align}
p_{\text{meas}} = (\vec{r}'_M)^T Y\, \vec{q}_s .
\end{align}
We can regard $Y$ as transforming the state, or alternatively as transforming the effect. In the latter case, we have $\vec{r}'_M \rightarrow Y^T \vec{r}'_M$. If we restrict attention to reversible transformations, then $Y\in \Omega^{\text{reversible}}$. Since the representation is orthogonal, $Y^T \in \Omega^{\text{reversible}}$ as well. 

\subsection{Qubit example}
To concretize the above discussion, it is instructive to examine the simplest nontrivial case: the qubit. The qubit provides a clear geometric picture of how reversible transformations act as rotations on the Bloch sphere, and how the unit effect remains invariant under these operations.

Consider the Pauli basis $\{ \mathbb{I}, \sigma_x, \sigma_y, \sigma_z \}$. Any qubit state $\rho$ can be expanded as
\begin{align}
\rho = \frac{1}{2} \left( \mathbb{I} + \vec{r} \cdot \vec{\sigma} \right),
\end{align}
where $\vec{\sigma} = (\sigma_x, \sigma_y, \sigma_z)$ and $\vec{r} = (r_x, r_y, r_z)$ is the Bloch vector satisfying $|\vec{r}| \leq 1$. We can represent $\rho$ as a vector in $\mathbb{R}^4$:
\begin{align}
\vec{q} = (1, r_x, r_y, r_z)^T,
\end{align}
where the first component corresponds to the identity operator. This representation allows effects and transformations to be seen as linear maps in a real vector space.

The \emph{unit effect} corresponds to the trace operation, which returns the normalisation of a quantum state. In the vector representation, it is
\begin{align}
\vec{I} = (1, 0, 0, 0)^T,
\end{align}
so that $\vec{I} \cdot \vec{q} = 1$, corresponding to $\mathrm{Tr}[\rho] = 1$.

Reversible transformations on the qubit correspond to unitary conjugations of the form
\begin{align}
\rho \mapsto U \rho U^\dagger,
\end{align}
for some unitary operator $U \in \mathrm{SU}(2)$. Such transformations preserve the purity of quantum states and correspond geometrically to rotations of the Bloch vector:
\begin{align}
\vec{r} \mapsto R \vec{r}, \qquad R \in \mathrm{SO}(3),
\end{align}
where $R$ is the adjoint representation of $U$. In the 4-dimensional real vector space, this transformation can be expressed as
\begin{align}
Y = 
\begin{pmatrix}
1 & 0 \\
0 & R
\end{pmatrix},
\qquad
\vec{q}' = Y \vec{q} = (1, R\vec{r}).
\end{align}
The block structure explicitly separates the identity component (which remains invariant) from the rotation acting on the Bloch vector.

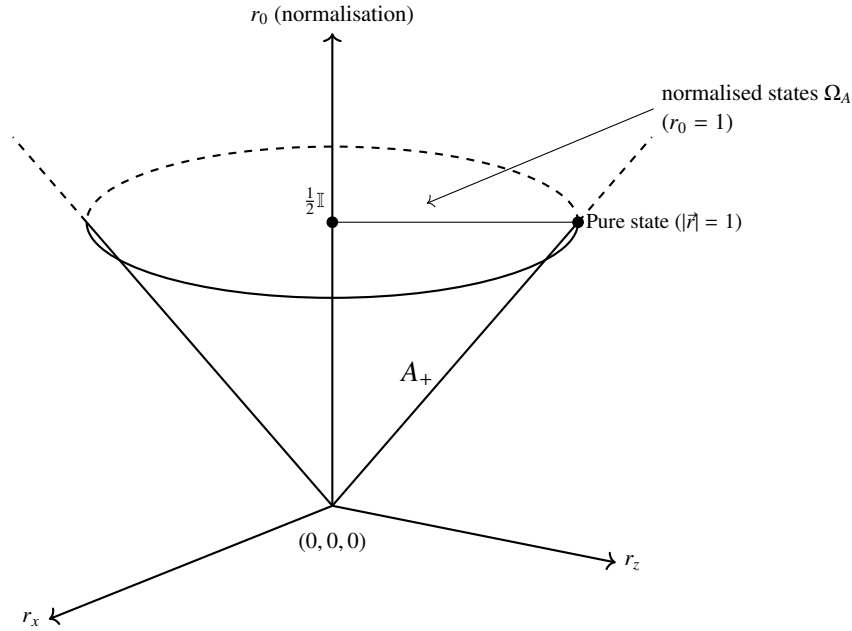
\begin{figure}[htbp]
    \centering
    \begin{tikzpicture}[scale=2.5]
        
        \def\h{1.5}  
        \def\w{1.3}  
        \def\d{0.4}  
        
        \draw[thick, ->] (0,0) -- (-1.5, -0.6) node[left] {$r_x$};
        \draw[thick, ->] (0,0) -- (1.5, -0.3) node[right] {$r_z$};
        
        \draw[thick, ->] (0,0) -- (0, 2.5) node[above] {$r_0$ (normalisation)};
        
        \draw[thick] (0,0) -- (-\w, \h);
        \draw[thick] (0,0) -- (\w, \h);
        
        \draw[thick, dashed] (\w,\h) arc (0:180:\w cm and \d cm);
        \draw[thick] (-\w,\h) arc (180:360:\w cm and \d cm);
        
        \draw[thick, dashed] (-\w, \h) -- (-1.3*\w, 1.3*\h);
        \draw[thick, dashed] (\w, \h) -- (1.3*\w, 1.3*\h);
        
        \draw[thin] (0,\h) -- (\w,\h);
        
        \filldraw[black] (0,\h) circle (0.8pt) node[above left, font=\small] {$\frac{1}{2}\mathbb{I}$};
        
        \filldraw[black] (\w,\h) circle (0.8pt) node[right, font=\small] {Pure state ($|\vec{r}| = 1$)};
        
        \node[font=\large] at (0.45, 0.7) {$A_+$};
        
        \draw[->, thin] (1.7, 2.1) node[right, align=left] {normalised states $\Omega_A$ \\ ($r_0 = 1$)} -- (0.5, 1.6);
        
        \node[below] at (0,-0.1) {$(0,0,0)$};
        
    \end{tikzpicture}
    \caption{{\bf The state cone $A_+$ for a qubit}. The depiction is  restricted to the $r_x$-$r_z$ plane for visual clarity (suppressing $r_y$). The vertical axis represents the normalisation component $r_0$. The normalised state space $\Omega_A$ is the horizontal slice at $r_0 = 1$, forming a solid disk (a cross-section of the Bloch sphere). Pure states lie on the boundary of this disk, while the maximally mixed state lies at the center. The cone extends infinitely upward, representing unnormalised states with $\mathrm{Tr}[\rho] > 1$.}
    \label{fig:qubit_cone}
\end{figure}

As a concrete illustration, consider the unitary $U = e^{-i \frac{\pi}{2} \sigma_z}$, corresponding to a rotation of $\pi$ around the $z$-axis. The associated rotation matrix is
\begin{align}
R = 
\begin{pmatrix}
-1 & 0 & 0 \\
0 & -1 & 0 \\
0 & 0 & 1
\end{pmatrix},
\end{align}
and the full transformation matrix $Y$ acting on $\vec{q}$ is
\begin{align}
Y = 
\begin{pmatrix}
1 & 0 & 0 & 0 \\
0 & -1 & 0 & 0 \\
0 & 0 & -1 & 0 \\
0 & 0 & 0 & 1
\end{pmatrix}.
\end{align}
The unit effect under such a reversible transformation satisfies
\begin{align}
Y^{\mathrm{T}} \vec{I} 
= 
Y^{\mathrm{T}} 
\begin{pmatrix}
1 \\ 0 \\ 0 \\ 0
\end{pmatrix}
= 
\begin{pmatrix}
1 \\ 0 \\ 0 \\ 0
\end{pmatrix}
= \vec{I}.
\end{align}
Thus, the unit effect $\vec{I}$ remains unchanged under any reversible transformation $Y$, reflecting the fact that unitary evolution preserves normalisation. Fig.~\ref{fig:qubit_cone} depicts a part of the state cone with the normalisation as one axis.

\section{Derivation of Quantum Theory from Axioms}

We now describe how standard
finite-dimensional quantum theory can be reconstructed from simple operational
principles within the generalised probabilistic theories (GPT) framework.
This reconstruction clarifies which
features of quantum theory are forced by general information-theoretic and
geometric requirements, and which features distinguish quantum theory from
other conceivable probabilistic theories. In particular, it shows how the
structure of the state space, the scaling of the number of degrees of freedom,
and the form of reversible dynamics can all be derived from a small set of
axioms.

We follow Hardy's axiomatic approach~\cite{Hardy}, formulated in terms of two basic
operational quantities: the dimension $N$, defined as the maximal number of
mutually distinguishable states in a single-shot measurement, and the number
of degrees of freedom $K$, defined as the number of real parameters required
to specify a state completely. The derivation clarifies how the functional
relation between $K$ and $N$ is compatible with physically natural axioms,
and how this already strongly constrains the possible theory.

\subsection{List of Axioms}

We now state the axioms following Hardy’s approach.

The \emph{number of degrees of freedom} $K$ is defined as the number of real parameters needed to completely specify the state of a system. The \emph{dimension} $N$ is defined as the maximum number of mutually distinguishable states in a single-shot measurement.

\begin{axiom}[Simplicity]\label{axi: simp}
The number of degrees of freedom $K$ is determined by a function of $N$ (i.e., $K=K(N)$ for $N = 1,2,\ldots$), and for any given $N$, $K(N)$ takes the minimum value consistent with the axioms.
\end{axiom}

\begin{axiom}[Subspaces]\label{axi: subspace}
A system whose state is constrained to belong to an $M$-dimensional subspace behaves like a system of dimension $M$.
\end{axiom}

\begin{axiom}[Composite systems]\label{axi: composite}
A composite system consisting of two subsystems $A$ and $B$ having dimensions $N_A$ and $N_B$ and degrees of freedom $K_A$ and $K_B$, respectively, has dimension $N=N_A N_B$ and number of degrees of freedom $K = K_A K_B$.
\end{axiom}

Remark: The Composite systems axiom is closely connected to the property that all states $\omega_{AB} \in \Omega_{AB}$ are uniquely determined by the joint statistics of all {\em local} measurements on the components, which is also known as \emph{tomographical locality}. Operational theories that are not tomographically local also exist~\cite{stueckelberg1960quantum}.

\begin{axiom}[Continuity]\label{axi: continu}
There exists a continuous reversible transformation on a system between any two pure states of the system.
\end{axiom}

The derivation of quantum theory in this framework is based on three main steps: 1) finding the relation $K=N^r$ where $r$ is a positive integer; 2) ruling out the classical case $K=N$; 3) constructing quantum theory for the remaining allowed case.
More specifically, first, the axioms imply
that the number of degrees of freedom must scale as
\(
K(N)=N^r
\)
for some positive integer $r$. Second, the case $r=1$, corresponding to
classical probability theory, is excluded by the continuity requirement on
reversible transformations. Third, the minimal remaining option,
\(
K=N^2,
\)
leads to the geometry of quantum state spaces. In the simplest nontrivial
case $N=2$, one obtains the Bloch ball together with continuous reversible
transformations acting as rotations, which is the characteristic qubit
structure. This provides the basic template for the full finite-dimensional
density-matrix formulation of quantum theory. We now describe the three-step derivation.

\subsection{From Axioms to the Functional Form $K=N^r$}

We now show that $K(N)$ must be of the form $K(N) = N^r$ for some positive integer $r$. (The argument is similar but not identical to Ref.\cite{Hardy}.) From Axiom~\ref{axi: subspace}, $K(N)$ must be a strictly increasing function of $N$. Furthermore, Axiom~\ref{axi: composite} requires that $K$ satisfies$$K(N_A N_B) = K(N_A) K(N_B),$$meaning $K$ is multiplicative. Based on these two properties, we establish the following proof.

\begin{proof}
Let $n \ge 2$ and $m \ge 2$ be any two integers. For any arbitrary positive integer $k$, there exists a non-negative integer $\ell$ such that
\begin{equation}
    n^\ell \le m^k < n^{\ell+1}.
\end{equation}
Taking the natural logarithm of the inequalities, we can bound the ratio of $\ell$ to $k$
\begin{equation}
    \ell \ln n \le k \ln m < (\ell+1) \ln n \implies \frac{\ell}{k} \le \frac{\ln m}{\ln n} < \frac{\ell+1}{k}.
\end{equation}
Since $K$ is a strictly increasing function and it is multiplicative, we have
\begin{equation}
    K(n)^\ell \le K(m)^k < K(n)^{\ell+1}.
\end{equation}
Taking the natural logarithm yields
\begin{equation}
    \ell \ln K(n) \le k \ln K(m) < (\ell+1) \ln K(n) \implies  \frac{\ell}{k} \le \frac{\ln K(m)}{\ln K(n)} < \frac{\ell+1}{k}.
\end{equation}

We now have two expressions bounded within an interval of size $\frac{1}{k}$. Therefore, the distance between the two ratios must be strictly less than $\frac{1}{k}$
\begin{equation}
    \left| \frac{\ln m}{\ln n} - \frac{\ln K(m)}{\ln K(n)} \right| < \frac{1}{k}.
\end{equation}

Since $k$ is an arbitrary positive integer, we can take the limit as $k \to \infty$. The term $\frac{1}{k} \to 0$, forcing the absolute difference to be zero
\begin{equation}
    \frac{\ln K(m)}{\ln K(n)} = \frac{\ln m}{\ln n} \implies \frac{\ln K(m)}{\ln m} = \frac{\ln K(n)}{\ln n}.
\end{equation}

Since this holds for any integers $m, n \ge 2$, the ratio must be a constant $\alpha$
\begin{equation}
    \frac{\ln K(N)}{\ln N} = \alpha \implies \ln K(N) = \alpha \ln N \implies K(N) = N^\alpha.
\end{equation}

Finally, because $K(N)$ must output a positive integer for any integer input $N$, and because $K(N)$ is strictly increasing, $\alpha$ must be a positive integer $r$. Therefore,
\begin{equation}
    K(N) = N^r, \quad r \in \mathbb{Z}^+.
\end{equation}
\end{proof}

\subsection{Ruling out the case $K=N$}

We now proceed to exclude the case $K(N)=N$ based on the Axioms.
Assume that $K=N$, so that the real vector space $A$ has dimension $N$, and the normalised state space $\Omega_A$ has affine dimension $N-1$.
By definition of the dimension $N$, there exists a set of $N$ perfectly distinguishable normalised states
$
\{\omega_1,\dots,\omega_N\} \subset \Omega_A,
$
together with effects $\{e_1,\dots,e_N\} \subset A^*$ such that
\begin{align}
e_i(\omega_j) = \delta_{ij},
\qquad
\sum_{i=1}^N e_i = u_A .
\end{align}
Since $\dim \Omega_A = N-1$, the states $\{\omega_i\}_{i=1}^N$ are affinely independent; equivalently, the $N-1$ vectors $\{\omega_i-\omega_1 \}^{N}_{i=2}$ are linearly independent. Consequently, every normalised state $\omega \in \Omega_A$ can be written uniquely as
\begin{align}
\omega = \sum_{i=1}^N p_i \omega_i,
\qquad
p_i \ge 0,
\qquad
\sum_{i=1}^N p_i = 1 .
\end{align}
Thus, $\Omega_A$ is the convex hull of $\{\omega_1,\dots,\omega_N\}$ and is therefore an $(N-1)$-dimensional simplex.
In a simplex, the extremal points coincide with its vertices. Hence, the pure states of the theory are exactly
$
\omega_1,\dots,\omega_N .
$
In particular, the set of pure states is finite and discrete.

Let $\mathcal T_A$ denote the group of reversible transformations. Each
$
T \in \mathcal T_A
$
is an invertible linear map on $A$ such that
\begin{align}
T(\Omega_A) = \Omega_A,
\qquad
u_A \circ T = u_A .
\end{align}
Since reversible transformations map pure states to pure states, every $T \in \mathcal T_A$ must permute the vertices of the simplex. Hence, the action of $\mathcal T_A$ on the set $\{\omega_1,\dots,\omega_N\}$ is given by permutations, and the group $\mathcal T_A$ is therefore finite (or at most discrete).

However, Axiom~\ref{axi: continu} (Continuity) requires that for any two pure states $\omega,\omega' \in \Omega_A$, there exists a continuous one-parameter family of reversible transformations
$
\{T_t\}_{t \in [0,1]} \subset \mathcal T_A
$
such that
\begin{align}
T_0 = \mathbb I,
\qquad
T_1(\omega) = \omega'.
\end{align}
Such a continuous path cannot exist if the pure state space is finite and discrete: continuous reversible transformations cannot interpolate between distinct vertices of a simplex without leaving the set of permutations.

In conclusion, the assumption $K=N$, corresponding to classical probability theory, is incompatible with Axiom~\ref{axi: continu}( Continuity). By Axiom~\ref{axi: simp}(Simplicity), the minimal remaining possibility is therefore
\begin{align}
K(N)=N^2.
\end{align}

\subsection{ Bloch Sphere for $N=2$}

We now consider the simplest nontrivial case, namely a system of dimension $ N=2$.
From the previous derivation, the number of degrees of freedom is $ K(2)=2^2=4$,
so the real vector space $A$ has dimension $4$. Imposing the normalisation
condition $u_A(\omega)=1$, the set of
normalised states $\Omega_A$ has dimension $3$.
The goal is to show that, under the axioms, $\Omega_A$ must be affinely
isomorphic to a three-dimensional Euclidean ball.

To characterise the set of states and the associated effects Hardy chooses fiducial states and effects together with an assumption that there is a classical bit as a subsystem.

Four fiducial effects are needed and two can be associated with a classical bit which is a subspace. Choose a set of fiducial effects $\{e_1,e_2,e_3,e_4\}$ that forms a
basis of the dual space $A^*$. By definition, the values
$
e_k(\omega),  k=1,\dots,4,
$
uniquely determine any state $\omega \in A$.
Assuming the existence of a classical bit subsystem, in line with Hardy's subspace axiom  we choose the fiducial effects such that two of them, say $e_1$ and $e_2$, are
pure effects that perfectly distinguish two pure states
$\omega_1,\omega_2 \in \Omega_A$, i.e.
\begin{equation}
e_i(\omega_j)=\delta_{ij}, i,j=1,2, \ \  \text{and} \ \ 
e_1 + e_2 = u_A .
\label{eq:classicalsubsystem}
\end{equation}
The two-dimensional face spanned by $\omega_1$ and $\omega_2$ behaves as a normalised classical subsystem. 
Furthermore, we choose the remaining fiducial states \(\omega_3, \omega_4\) to also be pure states, and choose the fiducial effects \(e_3, e_4\) to be the pure effects that perfectly identify them. Because of this choice, the effect \(e_k\) identifies the state \(\omega_k\):  
\begin{equation}
    e_k(\omega_k) = 1\, \text{for}\, k \in \{1, 2, 3, 4\}.
\label{eq:diagonal_ones}
\end{equation}\\

An important object is the bilinear pairing matrix between states and effects: 
\begin{align}
D_{ij} := e_i(\omega_j).
\end{align}
We can exploit the non-uniqueness of the fiducial elements to ensure $D$ is symmetric~\cite{Hardy}:
\begin{equation}
D_{ij}=D_{ji}
\label{eq:Dsymmetric}
\end{equation}
From Eq.~\ref{eq:classicalsubsystem}, Eq.~\ref{eq:diagonal_ones} and Eq.~\ref{eq:Dsymmetric},  $D$ takes the parametric form
\begin{align}
D = \begin{pmatrix} 1 & 0 & 1-a & 1-b \\ 0 & 1 & a & b \\ 1-a & a & 1 & c \\ 1-b & b & c & 1 \end{pmatrix},
\end{align}
for real parameters $a, b, c$. Characterising D is important because 
\begin{equation}
    e(\omega) = \left( \sum_i d_i e_i \right) \left( \sum_j c_j \omega_j \right) = \sum_{i,j} d_i c_j e_i(\omega_j) = \sum_{i,j} d_i D_{ij} c_j.
\end{equation}

The set of normalised states is a subset of a {\em three} dimensional real vector space. Let $\omega_0 \in \Omega_A$ be a fixed reference state. Every normalised state
$\omega \in \Omega_A$ can be written uniquely as
\begin{align}
\omega = \omega_0 + x,
\qquad
x \in V := \ker u_A .
\end{align}
Here, the kernel is defined as $\ker u_A = \{x \in A \mid u_A(x) = 0\}$, representing the subspace of zero-probability displacements between valid states.
Therefore, $\dim V = 3$. From this perspective, $\Omega_A$ can be viewed as a
compact convex subset of the three-dimensional real vector space $V$.

In our current fiducial basis, the unit effect is represented by the vector \(\vec{u} = (1, 1, 0, 0)^T\), as dictated by the relation \(e_1 + e_2 = u_A\). To explicitly separate the normalisation degree of freedom from the remaining degrees of freedom in \(V = \ker u_A\), we apply an invertible basis transformation matrix \(C\) such that the unit effect in the new basis becomes \((1, 0, 0, 0)^T\). This change of representation\footnote{In Dirac notation $C = (|1\rangle + |2\rangle)\langle 1| + |2\rangle\langle 2| + |3\rangle\langle 3| + |4\rangle\langle 4|$ and the inverse, under which the vectors transform is $C^{-1} = (|1\rangle - |2\rangle)\langle 1| + |2\rangle\langle 2| + |3\rangle\langle 3| + |4\rangle\langle 4|$ } transforms the pairing matrix \(D\) into a new matrix \(F\):
\begin{equation}
F=C^T D C=
\begin{pmatrix}
2 & 1 & 1 & 1\\
1 & 1 & a & b\\
1 & a & 1 & c\\
1 & b & c & 1
\end{pmatrix},
\end{equation}
This step isolates the normalisation coordinate and leaves three coordinates corresponding to the intrinsic degrees
of freedom in $V$.

After separating the normalisation direction $u_A$, the remaining three intrinsic coordinate $x$ directions are obtained by subtracting the uniform $\omega_0$ component $\frac{1}{2}(1,1,1,1)^T$ from the fiducial coordinate vectors. The Gram matrix of these intrinsic coordinates is given by
\begin{align}
G=
\begin{pmatrix}
\frac12 & a-\frac12 & b-\frac12\\
a-\frac12 & \frac12 & c-\frac12\\
b-\frac12 & c-\frac12 & \frac12
\end{pmatrix}.
\end{align}

The pure states are on a quadric surface in a three dimensional space associated with $G$. Pure states are extremal points of the compact convex set $\Omega_A$. Since continuous reversible transformations act transitively on pure states and preserve operational probabilities, they preserve an invariant inner product on $V$. This continuous symmetry forces the intrinsic component $x \in V$ of all pure states to lie on a single continuous orbit, uniquely satisfying a quadratic constraint.
Equivalently, there exists a symmetric bilinear form
$Q:V\times V\to\mathbb R$ such that pure states
$\omega=\omega_0+x$ are characterized by
\begin{align}
Q(x,x)=x^TG^{-1}x=\text{const}.
\end{align}
(The quadric surface arises because continuous reversible symmetry forces all pure states to have equal norm with respect to an invariant inner product, not simply because probabilities must lie in [0,1].) Thus, the set of pure states forms a two-dimensional quadric surface embedded in
the three-dimensional space $V$.

Finally, by an appropriate choice of fiducial coordinates, this geometry reduces to the Bloch ball.
Mathematically, the quadric surface defined by $G$ could be a hyperboloid or degenerate (if $G$ has zero eigenvalues). However, the strict positivity of effects ($0\le e(\omega)\le 1$) requires $\Omega_A$ to be a bounded, compact set. This operational requirement physically rules out unbounded geometries, demanding that the bilinear form $Q$ be strictly positive definite. The pure states must, therefore, form an ellipsoidal surface. Because any three-dimensional ellipsoid equipped with a continuous, transitively acting symmetry group is linearly isomorphic to a sphere,\footnote{An ellipsoid in $\mathbb{R}^3$ with a continuous group of symmetries acting transitively on its surface is equivalent under a linear change of basis to a ball with $\mathrm{SO}(3)$ symmetry. We now describe a systematic method for finding a basis with the desired properties. 1. Define a $G$-invariant (in this footnote $G$ refers to the symmetry group) inner product by Haar averaging: 
\begin{align}
    \langle v, w \rangle = \int_G (\rho(g)v) \cdot (\rho(g)w) \, dg = v^T A w, 
\end{align}
where $A = \int_G \rho(g)^T \rho(g) \, dg.$
2. Let $A = B^T B$ (Cholesky decomposition) and change coordinates via $\tilde{v} = Bv$. In these coordinates: (i) The inner product becomes the standard dot product: $\langle v, w \rangle = \tilde{v} \cdot \tilde{w}$ (ii) The representation $\tilde{\rho}(g) = B\rho(g)B^{-1}$ is orthogonal, (iii) The ellipsoid becomes the unit ball: $\{v : v^T A v = 1\} \to \{\tilde{v} : |\tilde{v}| = 1\}$. Since $G$ acts continuously and transitively on the sphere via orthogonal transformations, it must be exactly $\mathrm{SO}(3)$.} 
an appropriate change of basis on $V$ smoothly maps this geometry to the unit ball. In Hardy’s parametrization, choosing the canonical spherical representation corresponds to setting $a=b=c=\frac{1}{2}$, which simplifies the pairing matrix to
\begin{align}
D=
\begin{pmatrix}
1 & 0 & \tfrac12 & \tfrac12\\
0 & 1 & \tfrac12 & \tfrac12\\
\tfrac12 & \tfrac12 & 1 & \tfrac12\\
\tfrac12 & \tfrac12 & \tfrac12 & 1
\end{pmatrix}.
\end{align}
In this representation, $Q$ reduces to the standard Euclidean inner product on $V$. Thus, $\Omega_A$ is affinely isomorphic to the Euclidean unit ball, and its boundary sphere represents the pure states—rigorously recovering the Bloch-ball representation of qubit states.

The reversible quantum dynamics for N=2 are recovered.  
Reversible transformations preserve both the convex structure and the Euclidean inner product. Each transformation $T\in\mathcal T_A$ induces an element of the rotation group $\mathrm{SO}(3)$ acting on the Bloch sphere. In quantum theory, these transformations arise from the unitary conjugation of density operators. The group of special (determinant 1) unitary matrices $\mathrm{SU}(2)$ (any unitary matrix can be decomposed into a  special unitary matrix and a global phase) acts on states via
\begin{align}\omega \mapsto U \omega U^\dagger , \qquad U\in\mathrm{SU}(2),
\end{align}
inducing spatial rotations of the Bloch sphere through the adjoint action. This realizes $\mathrm{SU}(2)$ as a double cover of $\mathrm{SO}(3)$ ($\pm U\in$ SU(2) and both produce the same rotation). Thus, within the GPT framework, the emergence of the Bloch sphere and the continuity axiom naturally single out $\mathrm{SU}(2)$ as the Lie group implementing reversible dynamics for $N=2$.

To summarize the derivation, under Axioms \ref{axi: simp}--\ref{axi: continu} and the foundational geometric structure of GPTs, we have established two key properties of the state space:
\begin{itemize}
\item Degrees of Freedom: The number of parameters required to define a state scales as $K = N^2$.
\item The Qubit Structure: For the simplest non-trivial case of $N=2$, the normalised state space $\Omega_A$ emerges uniquely as a three-dimensional Bloch ball.
\end{itemize}

While this derivation rigorously recovers the standard density-matrix formulation for a two-level quantum system, extending these results to an arbitrary finite dimension $N > 2$ is substantially more mathematically involved. As initially observed in Hardy's foundational work, and systematically detailed in Müller's  modern review of probabilistic theories~\cite{muller2021probabilistic}, deriving the full complex Hilbert space formalism for higher dimensions cannot be achieved as directly as the $N=2$ case. Constructing the complete density-matrix state space and ensuring the reversible transformations correspond strictly to the projective unitary group $PU(N)$ requires invoking advanced technical machinery. For a rigorous generalisation to arbitrary $N$, readers are encouraged to consult the more advanced reconstructive literature. These extended proofs frequently rely on the algebraic classification of formally real Jordan algebras, specific higher-order interference postulates, or additional assumptions regarding the continuous reversibility of multi-partite systems.


\subsection{Further readings}
Hardy's 2001 framework initiated a foundational shift by demonstrating that quantum theory can be derived from operational and information-theoretic principles. Over the subsequent decade, multiple reconstructions established that quantum mechanics can be uniquely recovered from distinct sets of physical postulates. For example, Chiribella, D’Ariano, and Perinotti (2011)~\cite{chiribella2011} reconstructed the theory using six informational principles, culminating in the Purification Postulate. This principle dictates that every mixed state is the marginal of a pure state in a joint system, delineating a fundamental operational boundary between quantum and classical probability. Similarly, the approach of Masanes and Müller (2011)~\cite{masanes2011derivation} minimizes initial assumptions, recovering quantum theory solely through the self-consistency of state-space geometry, continuous reversible dynamics, and the formal structure of bipartite systems. Furthermore, Dakić and Brukner (2011)~\cite{dakic2011quantum} demonstrated that the theory emerges from just three postulates: a bounded information capacity, continuous reversible evolution, and the requirement of local tomography for composite systems. Complementing these approaches, which rely heavily on structural properties of composite systems, Barnum, Müller, and Ududec (2014)~\cite{barnum2014higher} showed that state-space geometry can be rigorously constrained by single-system postulates. In particular, they demonstrated that the absence of higher-order interference, together with continuous reversible dynamics and the existence of projective measurements, strongly restricts the admissible operational framework, bringing it close to that of standard quantum mechanics.


\section{Nonlocality in GPTs}

The framework of Generalised Probabilistic Theories (GPTs) allows us to view classical, quantum mechanics, and post-quantum theories not as isolated theories, but as specific instances within a broader landscape of possible physical laws. Within this framework, the phenomenon of nonlocality acts as a crucial discriminator, distinguishing these theories based on the strength of correlations they allow.

To navigate this landscape, we first introduce the necessary mathematical tools for quantifying nonlocality. We begin by revisiting Bell's argument and deriving the CHSH inequality, which serve as the fundamental metrics for correlation strength. We then discuss Tsirelson's bound, which identifies the precise limit imposed by quantum mechanics on these metrics.

Equipped with these analytical tools, we then enter the ``Box World" of GPTs. Here, both algebraic constraints and geometric representations are employed to map out the spectrum of physical theories. These regions, including classical, quantum, and post-quantum, are distinguished by the strength of correlations they admit. Specifically, we illustrate how classical and quantum sets are situated within the geometric framework of the Gbit, the non-signalling polytope and convex compact bodies, extending to the maximum nonlocality exhibited by the PR box.

\subsection{Bell non-locality}


In this section we follow Bell's analysis of the EPR argument to show why no local hidden-variable theory can reproduce the statistical predictions of quantum mechanics~\cite{Bell_speak}. We begin by introducing the framework of local hidden-variable models and illustrating it with Bell's example. We then show that any such model must exhibit a characteristic 
non-smoothness in its correlation function, and derive Bell's inequality as the rigorous expression of this constraint. This inequality is then 
reformulated in the CHSH form, which is more amenable to experimental test, and we discuss the Tsirelson bound that separates quantum from classical correlations. We close with a geometric picture that recasts the CHSH inequality in terms of vector inner products, offering an additional perspective on the structure of the bound.

\subsubsection{Local hidden variable models (LHV) and Bell's example}
Consider a low–energy proton–proton scattering process in which the two outgoing 
protons are produced in a spin–singlet state. Each proton travels toward a distant 
detector equipped with a spin filter. If both filters are oriented along the same axis, 
the outcomes exhibit perfect anti–correlation: whenever detector~1 records spin 
up, detector~2 invariably records spin down. The crucial point is that the 
measurement at detector~1 seems to determine instantaneously the outcome at 
detector~2, even though the second proton may be spatially remote. This is the 
central tension raised by EPR~\cite{BELL1964}: the apparent conflict between quantum predictions 
and the principle of locality.

Bell notes that one might suppose that quantum correlations originate from 
hidden variables that predetermine the outcomes. This view relies on two 
fundamental assumptions: \textbf{Locality}, meaning that measurement choices at 
one detector cannot instantaneously influence outcomes at the distant detector, 
and \textbf{Realism}, meaning that outcomes are predetermined by hidden 
properties. Together, these constitute the hypothesis of \textbf{local realism}.

This idea can be formalized by introducing a \textbf{local hidden-variable 
model}. In such a model, the measurement outcomes are determined by the local 
measurement settings and by a hidden variable \(\bm{\lambda}\), which represents the 
shared past variables that may influence both particles. Thus, for local 
settings \(a\) and \(b\), the outcomes take the form
\begin{equation}
A=A(a,\bm{\lambda}), \quad B=B(b,\bm{\lambda}),
\end{equation}
where \(A,B=\pm1\). The locality condition is expressed by the absence of any 
dependence of \(A\) on the distant setting \(b\), and of \(B\) on the distant 
setting \(a\).

Bell gives a simple example of such a local hidden-variable model~\cite{BELL1964}. Let
\(\bm{\lambda}\) be a unit vector uniformly distributed over the unit sphere, and
define
\begin{equation}
A(a,\bm{\lambda})=\mathrm{sign}(a\cdot\bm{\lambda}),\qquad
B(b,\bm{\lambda})=-\mathrm{sign}(b\cdot\bm{\lambda}).
\end{equation}
This model is local because each outcome depends only on the local measurement
setting and on the shared hidden variable \(\bm{\lambda}\). When \(a=b\), the two
outcomes are perfectly anti-correlated, since
\begin{equation}
A(a,\bm{\lambda})B(a,\bm{\lambda})=-1 .
\end{equation}

\begin{figure}[htbp]
    \centering
    \begin{tikzpicture}[scale=1.35, >=Stealth]
        
        \def\angleA{90}
        \def\angleB{50}
        \def\r{2.4}
        
        \fill[gray!25, opacity=0.65] 
            (0,0) -- (180:\r) arc (180:140:\r) -- cycle;

        \fill[gray!25, opacity=0.65] 
            (0,0) -- (0:\r) arc (0:-40:\r) -- cycle;

        \draw[gray!60] (180:\r) arc (180:140:\r);
        \draw[gray!60] (0:\r) arc (0:-40:\r);

        \draw[thick] (-2.9,0) -- (2.9,0);
        \draw[thick, dashed, gray!80] (140:2.9) -- (-40:2.9);

        \draw[->, thick] (0,0) -- (\angleA:2.2) node[above] {$\vec a$};
        \draw[->, thick] (0,0) -- (\angleB:2.2) node[above right] {$\vec b$};

        \draw[->] (90:0.75) arc (90:50:0.75);
        \node at (68:0.95) {$\theta$};

        \draw[->, very thick, red!80!black] (0,0) -- (-22:2.1)
            node[pos=0.68, below] {$\vec{\lambda}$};

        \node[font=\small] at (155:1.8) {special zone};

    \end{tikzpicture}

    \caption{ {\bf{Bell's hidden variable model}}. $A(\vec{a}, \vec{\lambda}) = \operatorname{sign}(\vec{a} \cdot \vec{\lambda}) ; \quad B(\vec{b}, \vec{\lambda}) = -\operatorname{sign}(\vec{b} \cdot \vec{\lambda})$. $A = -B$ unless $\vec{\lambda}$ in special zone, which happens with probability $\frac{2\theta}{2\pi} = \frac{\theta}{\pi}$.}
    \label{fig:bells_hidden_variable}
\end{figure}
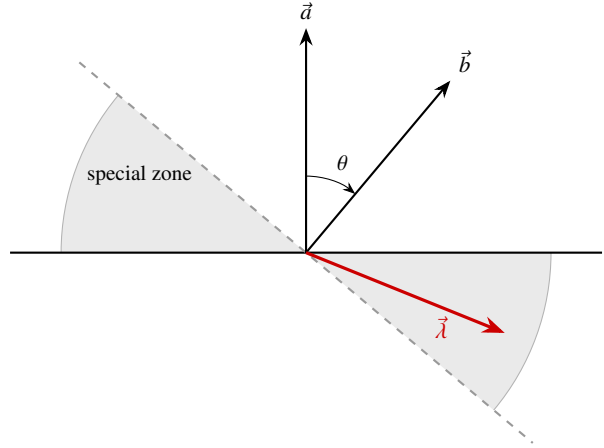

More generally, let \(\theta\) be the angle between \(a\) and \(b\), as in~\autoref{fig:bells_hidden_variable}. In the overlapping region where \(a\cdot\bm{\lambda}\) and \(b\cdot\bm{\lambda}\) are both positive or both negative, the two quantities have the same sign, and hence \(AB=-1\). They differ only when \(\bm{\lambda}\) lies in the angular region between the two sign-changing boundaries determined by \(a\) and
\(b\). Since \(\bm{\lambda}\) is uniformly distributed, the probability of this
region is \(\theta/\pi\). Therefore
\begin{equation}
\begin{aligned}
P_{\mathrm{LHV}}(a,b)
&=\langle A(a,\bm{\lambda})B(b,\bm{\lambda})\rangle_{\bm{\lambda}}  \\
&=\left(1-\frac{\theta}{\pi}\right)(-1)
+\frac{\theta}{\pi}(+1)  \\
&=-1+\frac{2\theta}{\pi}.
\end{aligned}
\end{equation}

This linear dependence on $\theta$ stands in sharp contrast to the quantum 
prediction $P_{\mathrm{QM}}=-\cos\theta$, which is smooth and has a vanishing 
derivative at $\theta=0$. One might ask whether a more sophisticated local hidden-variable model could reproduce the smooth quantum curve. The next subsection explains why this is not possible.

\subsubsection{Bell's inequality showing all LHVs unsmooth}
Bell's inequality proves that under locality and perfect anti-correlation, any LHV model has an unsmooth kink, satisfying a linear bound on the difference between correlations measured at different settings.

The setup is the same as before: in a local
hidden-variable model, the measurement outcomes are predetermined by a
hidden variable $\bm{\lambda}$,
\begin{equation}
    A( a,\bm{\lambda})=\pm 1,
    \qquad
    B( b,\bm{\lambda})=\pm 1,
\end{equation}
and the correlation function is
\begin{equation}
    P( a, b)
    =
    \int d\bm{\lambda}\,\rho(\bm{\lambda})
    A( a,\bm{\lambda})B( b,\bm{\lambda}),
    \label{eq:lhv_correlation}
\end{equation}
where $\rho(\bm{\lambda})$ is a normalised probability distribution.
Imposing perfect anti-correlation for identical settings $A(a,\bm{\lambda})=-B( a,\bm{\lambda})$,
for all $\bm{\lambda}$, allows Eq.~\eqref{eq:lhv_correlation} to be
rewritten as
\begin{equation}
    P( a, b)
    =
    -\int d\bm{\lambda}\,\rho(\bm{\lambda})
    A(a,\bm{\lambda})A( b,\bm{\lambda}).
    \label{eq:lhv_correlation_aa}
\end{equation}

Now introduce a third measurement direction $ c$. From
Eq.~\eqref{eq:lhv_correlation_aa},
\begin{align}
    P(a, b)-P(a, c)
    &=
    -\int d\bm{\lambda}\,\rho(\bm{\lambda})
    A( a,\bm{\lambda})
    \left[
        A(b,\bm{\lambda})-A(c,\bm{\lambda})
    \right].
\end{align}
Since $A(b,\bm{\lambda})^2=1$, this can be written as
\begin{align}
    P(a,b)-P(a,c)
    &=
    -\int d\bm{\lambda}\,\rho(\bm{\lambda})
    A(a,\bm{\lambda})A(b,\bm{\lambda})
    \left[
        1-A(b,\bm{\lambda})A(c,\bm{\lambda})
    \right].
\end{align}
Taking the absolute value and using
$|A(a,\bm{\lambda})A(b,\bm{\lambda})|=1$, we obtain
\begin{align}
    \left|
    P(a,b)-P(a,c)
    \right|
    &\leq
    \int d\bm{\lambda}\,\rho(\bm{\lambda})
    \left[
        1-A(b,\bm{\lambda})A(c,\bm{\lambda})
    \right]
    =
    1+P(b,c).
\end{align}
Therefore every local hidden-variable model satisfying perfect
anti-correlation must obey Bell's original inequality
\begin{equation}
    1+P(b,c)
    \geq
    \left|
    P(a,b)-P(a,c)
    \right|.
    \label{eq:bell_original_inequality}
\end{equation}

\begin{figure}[htbp]
    \centering
    \begin{tikzpicture}[scale=3, >=stealth]

        \draw[->, thick] (-0.05, -1) -- (2.5, -1) node[right] {Angle};
        \draw[->, thick] (0, -1.2) -- (0, 0.5) node[above] {\(P\)};

        \draw (1, -1) +(0, 0.03) -- +(0, -0.03) node[below=3pt] {\(\theta\)};
        \draw (2, -1) +(0, 0.03) -- +(0, -0.03) node[below=3pt] {\(2\theta\)};
        \node[left=3pt] at (0.0, -1) {\(-1\)};

        \coordinate (P0) at (0, -1);
        \coordinate (P1) at (1, -0.7);
        \coordinate (P2) at (2, 0.2);

        \draw[thick, domain=0:2.15, samples=50] plot (\x, {-1 + 0.3*\x*\x}) 
            node[right] {QM Curve};

        \draw[dashed, thick] (P0) -- (P1) node[midway, above left=-2pt] {\(m_1\)};
        \draw[dashed, thick] (P1) -- (P2) node[midway, above left=-2pt] {\(m_2\)};

        \draw[dotted, thick] (0, -1) -- (2.15, -0.355) 
            node[right, align=left] {LHV Bound\\\(m_2 \le m_1\)};

        \fill (P0) circle (0.7pt);
        \fill (P1) circle (0.7pt);
        \fill (P2) circle (0.7pt);


    \end{tikzpicture}
    \caption{ {\bf Geometric visualization of Bell's inequality near \(\theta = 0\)}. The local hidden variable (LHV) bound requires that the slope of the second secant line (\(m_2\)) is not greater than the slope of the first (\(m_1\)). Because the horizontal steps are equal (\(\theta\)), comparing the vertical rises gives the inequality \(P(2\theta) - P(\theta) \leq P(\theta) - (-1)\). Quantum mechanics predicts a smooth curve that bends upwards (\(m_2 > m_1\)), violating this linear bound.}

    \label{fig:local_quantum}
\end{figure}
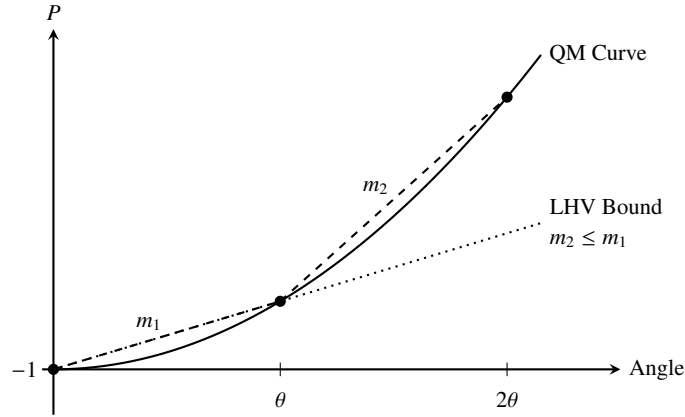

Let \(b\) serve as a fixed reference setting, and denote by \(\theta\) the relative angle between Alice's setting and this reference. We compare two nearby
Alice settings \(a\) and \(a'\) such that
$a-b=\theta$, $a'-b=2\theta$ and $a'-a=\theta$. With this notation, the preceding sign-flip picture leads to
\begin{equation}
    P(2\theta)-P(\theta) \leq P(\theta)-P(0),
\end{equation}
or equivalently, since $P(0)=-1$,
\begin{equation}
\label{eq:theta}
    P(2\theta)-P(\theta) \leq P(\theta)+1 .
\end{equation}
Because the two horizontal intervals in Eq.~(\ref{eq:theta}) are both \(\theta\), this inequality compares 
only the two vertical rises. It requires that the rise from \(\theta\) to \(2\theta\) is not larger than the rise from \(0\) to \(\theta\). Hence the deviation from perfect anti-correlation can grow at most linearly as the angle is opened from \(\theta=0\). This can be seen clearly in~\autoref{fig:local_quantum}. Therefore, as the measurement angle \(a'\) is varied across \(a\), or equivalently as \(\theta\) passes through zero, the local hidden-variable correlation departs from \(P=-1\) linearly on both sides of \(\theta=0\). 
For nonzero \(\rho\), the two sides have different slopes. This produces a \emph{kink} at \(\theta=0\), so the derivative does not exist there, and the correlation function is not stationary at its minimum value \(P(0)=-1\). Consequently, the correlation function cannot vary smoothly with the measurement setting \(a\).


In contrast, the quantum prediction
\begin{equation}
P_{\mathrm{QM}}(a,b) = -\cos\theta
\end{equation}
is a smooth function of $\theta$ and has a vanishing derivative at $\theta=0$. This contrast in smoothness highlights one of Bell's key insights: no local hidden-variable
model can generically reproduce the smooth angular dependence of quantum correlations.

\subsubsection{From Bell's inequality to the CHSH inequality}

Bell's original inequality~\eqref{eq:bell_original_inequality} establishes the essential constraint on local hidden-variable theories, but involves three measurement directions in a combination that is not straightforward to optimise experimentally. Specifically, the three-direction setup (e.g., $a$, $b$, and $c$) requires Alice and Bob to ``share'' measurement axes to evaluate the necessary correlations. For instance, Alice might measure along $a$ while Bob measures along $b$ in one run, and later Alice measures along $b$ while Bob measures along $c$. This implies that both observers must be capable of measuring along the exact same spatial axis (such as $b$), imposing a highly stringent practical requirement that their remote measurement devices have perfectly aligned spatial coordinate systems.

Building on Bell's observation, Clauser, Horne, Shimony and Holt (CHSH)~\cite{CHSH} constructed an equivalent formulation using two settings on each side, $(a,a')$ for Alice and $(b,b')$ for Bob, 
which is more directly accessible to experiment. This four-direction approach allows for a completely independent choice of measurement settings, decoupling Alice and Bob and eliminating the need for perfectly aligned shared reference frames. Their key idea was to form 
an algebraic identity that holds for every hidden variable~$\bm{\lambda}$:
\begin{equation}
C(\bm{\lambda})
= A(a,\bm{\lambda})B(b,\bm{\lambda})
  +A(a,\bm{\lambda})B(b',\bm{\lambda})
  +A(a',\bm{\lambda})B(b,\bm{\lambda})
  -A(a',\bm{\lambda})B(b',\bm{\lambda}) ,
\end{equation}
which can be factored as $A(a,\bm{\lambda})[B(b,\bm{\lambda})+B(b',\bm{\lambda})] + A(a',\bm{\lambda})[B(b,\bm{\lambda})-B(b',\bm{\lambda})]$. Since $A,B \in \{\pm 1\}$, one of the terms in the brackets is $\pm 2$ and the other is $0$, implying $|C(\bm{\lambda})| = 2$. Averaging over $\bm{\lambda}$ with distribution $\rho(\bm{\lambda})$ then yields the CHSH inequality:
\begin{equation}
|P(a,b) + P(a,b') + P(a',b) - P(a',b')| \le 2,
\end{equation}
satisfied by all local hidden–variable theories. Quantum mechanics predicts
values up to $2\sqrt{2}$ (the Tsirelson bound discussed later) for suitable choices of $(a,a',b,b')$, thereby violating
this bound. The CHSH result thus reformulates Bell's locality constraint in a form directly suited for laboratory tests.

Bell’s analysis transforms the EPR paradox from a philosophical debate into a concrete experimental question: does nature respect the constraints of locality, or do quantum correlations truly exceed the bounds imposed by any local 
hidden–variable model? The CHSH inequality provides a clear quantitative test. 
If experiments yield violations consistent with the quantum prediction, then 
no theory that maintains local realism can reproduce the observed correlations. 
Bell emphasized that this conclusion does not rely on the detailed structure of 
quantum theory, but follows directly from the minimal assumptions of locality 
and predetermined outcomes.

\subsubsection{Tsirelson's bound for the CHSH inequality}

While the CHSH inequality bounds all local hidden-variable theories by 2, quantum mechanics allows for violations of this limit. However, quantum nonlocality is not unbounded. The maximal possible value for the CHSH observable within the framework of quantum mechanics is limited to $2\sqrt{2}$. The origin of this ceiling lies in the structural constraints of quantum theory.


In a foundational 1980 paper, Tsirelson studied how algebraic constraints on observables, such as Hermiticity, bounded spectra, and commutativity, restrict the set of achievable bipartite correlations~\cite{Tsirelson1980}. These constraints can be formulated in the language of \(C^*\)-algebras, which provide an abstract framework for observables and their relations. In this setting, quantum theory appears as a particular realization of these structural constraints, leading to the fundamental limit now known as the Tsirelson bound.

A \(C^*\)-algebra is an algebra equipped with a product, an adjoint operation \(A\mapsto A^*\), and a norm satisfying the \(C^*\)-identity \(\|A^*A\|=\|A\|^2\). In this framework, Hermitian elements \(A=A^*\) represent observables, positive elements represent physically allowed positive quantities, and states are positive normalised linear functionals assigning expectation values to observables. In finite-dimensional quantum theory, this abstract description reduces to the usual matrix formulation. For example, a qubit is described by the algebra \(M_2(\mathbb C)\) of all \(2\times2\) complex matrices, whose Hermitian elements are observables. A density matrix \(\rho\) defines a state functional by
\begin{equation}
    \omega_\rho(A)=\operatorname{Tr}(\rho A),
\end{equation}
for any observable \(A\in M_2(\mathbb C)\). In the Bloch representation,
\begin{equation}
    \rho=\frac{1}{2}\left(\mathbb I+r_xX+r_yY+r_zZ\right),
    \qquad |{\bf r}|\leq 1,
\end{equation}
where the Bloch components are the expectation values of the Pauli observables,
\begin{equation}
    r_x=\omega_\rho(X),\qquad
    r_y=\omega_\rho(Y),\qquad
    r_z=\omega_\rho(Z).
\end{equation}
Thus, in finite dimensions, the density-matrix description is equivalently a
state-functional description on a \(C^*\)-algebra.

From the viewpoint of GPTs, a \(C^*\)-algebraic theory therefore gives a specific state-effect model: states are positive normalised functionals and effects are elements \(e\) satisfying \(0\leq e\leq \mathbb I\)~\cite{VanDeWetering2019EffectTheoretic}. 
Since GPTs provide a broader operational framework containing finite-dimensional classical and quantum theories as special cases~\cite{plavala2023general}, \(C^*\)-algebraic classical and quantum theories can be viewed as structured examples within the GPT framework, rather than as the most general GPTs.

Consider a scenario with observables \( A_k \) and \( B_l \) for \( k = 1, \dots, m \) and \( l = 1, \dots, n \), where each \( A_k \) commutes with every \( B_l \). Tsirelson investigates how algebraic and geometric constraints on such observables determine which sets of expectation values \( c_{kl} = \langle A_k B_l \rangle \) can arise within quantum theory. His analysis connects operator-algebraic, Hilbert space, and geometric formulations of bipartite correlations. One such formulation is given by the following theorem, which establishes the equivalence of four conditions for real numbers \( c_{kl} \):

\begin{enumerate}
    \item There exists a \( C^* \)-algebra \( \mathcal{A} \) with identity \(\mathbb I\),(a complex algebra equipped with an adjoint operation and a compatible norm, commonly used as an abstract framework for observables), together with Hermitian elements \( A_k, B_l \in \mathcal{A} \), and a state \( f \) on \( \mathcal{A} \) (a normalised positive linear functional assigning expectation values to observables) such that:

    \begin{equation}
    A_k B_l = B_l A_k, \quad -\mathbb{I} \leq A_k \leq \mathbb{I}, \quad -\mathbb{I} \leq B_l \leq \mathbb{I}, \quad f(A_k B_l) = c_{kl}.
    \end{equation}
    It characterizes correlations that are compatible with a broad class of physical models, including both classical and quantum systems, under the assumption that observables from distinct subsystems commute. This assumption aligns with the structure of tensor-product quantum systems and is sufficient to ensure no-signalling, though the theorem itself does not treat no-signalling as a primitive constraint. The requirement that all observables be Hermitian and bounded by the identity reflects physical plausibility, particularly in the context Tsirelson emphasizes: a system consisting of two spin-\(\tfrac{1}{2}\) particles, where measurement outcomes are restricted to \(\pm 1\).

    \item There exist Hermitian operators \( A_k, B_l \) on a Hilbert space \( \mathcal{H} \), and a density matrix \( W \) (positive semidefinite with unit trace), such that:
    \begin{equation}
    \text{spec}(A_k), \text{spec}(B_l) \subseteq [-1, +1], \quad \text{Tr}(A_k B_l W) = c_{kl}.
    \end{equation}

    This condition corresponds to the standard formulation of quantum theory, in which correlations are realized as expectation values of Hermitian operators acting on a Hilbert space. It directly connects to the physical picture of measurements on entangled states and shows that the correlation matrix \( \{ c_{kl} \} \) must lie within the set allowed by quantum mechanics—constrained both by the structure of the state space and the operator norms. In this formulation, realizability is grounded in the existence of a density matrix and bounded-spectrum observables.

    \item Same as (2), but additionally the operators satisfy:
    \begin{equation}
    A_k^2 = \mathbb{I}, \quad B_l^2 = \mathbb{I}, \quad \text{Tr}(A_k W) = \text{Tr}(B_l W) = 0,
    \end{equation}
    and the Hilbert space factorizes as \( \mathcal{H} = \mathcal{H}_1 \otimes \mathcal{H}_2 \), with \( A_k = A_k^{(1)} \otimes \mathbb{I}^{(2)} \), \( B_l = \mathbb{I}^{(1)} \otimes B_l^{(2)} \), where \( A_k^{(1)} \), \( B_l^{(2)} \) act on their respective subsystems, and the anticommutators of operators are proportional to identity operators. The dimensions of the Hilbert spaces obey:
    \begin{equation}
    \begin{aligned}
    2 \log_2 \dim \mathcal{H}_1 &\leq 
        \begin{cases}
        m & \text{if } m \text{ even} \\
        m + 1 & \text{if } m \text{ odd}
        \end{cases}, \\
    2 \log_2 \dim \mathcal{H}_2 &\leq 
        \begin{cases}
        n & \text{if } n \text{ even} \\
        n + 1 & \text{if } n \text{ odd}
        \end{cases}.
    \end{aligned}
    \end{equation}

    This is the most physically grounded and structured of the four formulations—closely matching real-world models of entangled spin-\(\tfrac{1}{2}\) particles. The additional constraints, including eigenvalues restricted to \( \pm 1 \), tensor-product structure, and bounded Hilbert space dimensions, ensure that this setting corresponds to what is operationally realizable in finite-dimensional quantum systems. It also serves as the standard framework in which maximal quantum violations, such as the Tsirelson bound in the CHSH scenario, can be explicitly achieved.

    \item There exist unit vectors \( x_k, y_l \) in a Euclidean space of dimension \( m+n \) such that:
    \begin{equation}
    \langle x_k, y_l \rangle = c_{kl}.
    \end{equation}
   This geometric formulation expresses quantum correlations as inner products between unit vectors in Euclidean space. It provides a conceptually simple perspective, reducing the analysis of such correlations to the geometry of vector configurations. In particular, it allows a direct derivation of the Tsirelson bound in the CHSH scenario, where the sum of selected inner products is maximized when the vectors are optimally oriented.
\end{enumerate}

This last, geometric formulation has important implications. Since the inner product of unit vectors is bounded by 1 in magnitude, it follows that quantum correlations are constrained by geometry. In the special case $m = n = 2$, this framework yields Tsirelson's bound.

To see this, let us define observables \( A_1, A_2, B_1, B_2 \) corresponding to spin measurements along different directions on a pair of entangled spin-\(\frac{1}{2}\) particles. With a suitable choice of measurement axes and a maximally entangled state \( W \), the following linear combination of correlators:
\begin{equation}
S = \langle A_1 B_1 \rangle + \langle A_1 B_2 \rangle + \langle A_2 B_1 \rangle - \langle A_2 B_2 \rangle
\end{equation}
can be written in quantum form as:
\begin{equation}
S = \operatorname{Tr}\left[(A_1 B_1 + A_1 B_2 + A_2 B_1 - A_2 B_2) W\right].
\end{equation}
By choosing observables with eigenvalues in \( [-1,1] \) and anticommutation relations mimicking spin-\(\frac{1}{2}\) operators, Tsirelson showed that this quantity can attain its quantum maximum:
\begin{equation}
S = 2\sqrt{2}.
\end{equation}

An explicit algebraic upper bound can be derived as follows. By a series of exact manipulations, the operator combination can be rewritten as:
\begin{align}
A_1 B_1 + A_1 B_2 + A_2 B_1 - A_2 B_2 &= \frac{1}{\sqrt{2}}(A_1^2 + A_2^2 + B_1^2 + B_2^2) \nonumber \\
&\quad - \frac{\sqrt{2}-1}{8}((\sqrt{2}+1)(A_1 - B_1) + A_2 - B_2)^2 \nonumber \\
&\quad - \frac{\sqrt{2}-1}{8}((\sqrt{2}+1)(A_1 - B_2) - A_2 - B_1)^2 \nonumber \\
&\quad - \frac{\sqrt{2}-1}{8}((\sqrt{2}+1)(A_2 - B_1) + A_1 + B_2)^2 \nonumber \\
&\quad - \frac{\sqrt{2}-1}{8}((\sqrt{2}+1)(A_2 + B_2) - A_1 - B_1)^2.
\end{align}
Each squared term on the right-hand side is positive semidefinite. Since all observables satisfy \( A_k^2 = B_l^2 = \mathbb{I} \), we conclude:
\begin{equation}
A_1 B_1 + A_1 B_2 + A_2 B_1 - A_2 B_2 \leq \frac{1}{\sqrt{2}}(4 \cdot \mathbb{I}) = 2\sqrt{2} \cdot \mathbb{I},
\end{equation}
and hence,
\begin{equation}
S = \operatorname{Tr}[(A_1 B_1 + A_1 B_2 + A_2 B_1 - A_2 B_2) W] \leq 2\sqrt{2}.
\end{equation}
This establishes the Tsirelson bound as a rigorous upper limit for quantum correlations in the CHSH scenario.

The fundamental difference between quantum and classical bounds arises from the mathematical structure of the correlation functions. In classical local hidden-variable theories, the observables can be represented within a commutative algebra. That is, \([A_i,A_j]=0\), \([B_i,B_j]=0\), and \([A_i,B_j]=0\) for all \( i, j \). The correlations can therefore be understood as averages over classical
joint probability distributions. In this setting, operator expressions obey familiar algebraic identities such as:
\begin{equation}
|A_1 B_1 + A_1 B_2 + A_2 B_1 - A_2 B_2| \leq |B_1 + B_2| + |B_1 - B_2| \leq 2.
\end{equation}
As a result, the CHSH combination is bounded by \( |S| \leq 2 \). This constraint arises purely from the commutativity of the observables and the deterministic structure of classical models.

\subsubsection{CHSH inequalities as vector inner products}

Consider the standard CHSH-type Bell operator
\begin{equation}
\mathcal{B} = A B + A B' + A' B - A' B',
\end{equation}
which is a Hermitian observable, i.e., $\mathcal{B}^\dagger = \mathcal{B}$.
To evaluate its maximal expectation value, one typically chooses the local observables as $A=Z$, $B=\frac{X+Z}{\sqrt{2}}$, $A'=X$, $B'=\frac{Z-X}{\sqrt{2}}$, where \(X\) and \(Z\) are the Pauli operators acting on each subsystem. The CHSH operator can be simplified as
\begin{equation}
\mathcal{B}
= A \otimes (B + B') + A' \otimes (B - B')
= \sqrt{2} \, (X \otimes X + Z \otimes Z).
\end{equation}
This operator acts on the two-qubit Hilbert space and attains its maximal quantum value
$|\langle \mathcal{B}\rangle| = 2\sqrt{2}$ for maximally entangled states.

\vspace{1em}
\noindent
The four Bell states form an orthonormal basis:
\begin{align}
|\Phi^\pm\rangle &= \frac{1}{\sqrt{2}}\big(|00\rangle \pm |11\rangle\big), \label{eq:bell-phi}\\
|\Psi^\pm\rangle &= \frac{1}{\sqrt{2}}\big(|01\rangle \pm |10\rangle\big). \label{eq:bell-psi}
\end{align}

\begin{align}
\rho_{\Phi^+}=|\Phi^+\rangle\langle\Phi^+| &= \frac{1}{4}\big(
I\otimes I
+ X\otimes X
- Y\otimes Y
+ Z\otimes Z
\big), \label{eq:phi-plus-pauli}\\[4pt]
\rho_{\Psi^-}=|\Psi^-\rangle\langle\Psi^-| &= \frac{1}{4}\big(
I\otimes I
- X\otimes X
- Y\otimes Y
- Z\otimes Z
\big). \label{eq:psi-minus-pauli}
\end{align}

Hence $\mathcal{B}$ is supported on the two-dimensional subspace
$\mathrm{span}\{|\Phi^+\rangle,|\Psi^-\rangle\}$ and can be written as
\begin{equation}
\mathcal{B} =
2\sqrt{2}\!\left(
|\Phi^+\rangle\!\langle\Phi^+|
-|\Psi^-\rangle\!\langle\Psi^-|
\right)
\end{equation}
and suppose we restrict it to the space spanned by $\rho_{\Phi^+}$ and $\rho_{\Psi^-}$, or in other words, $\pi_1=|\Phi^+\rangle\langle\Phi^+|$ and $\pi_2=|\Psi^-\rangle\langle\Psi^-|$, the operator can be expressed as
\begin{equation}
\vec{B}
= \begin{bmatrix}
2\sqrt{2}\\[4pt]
-2\sqrt{2}
\end{bmatrix}.
\end{equation}
Geometrically, this \(2\times1\) vector represents \(\mathcal{B}\) restricted to
the Bell-projector sector generated by \(\pi_{1}\) and \(\pi_{2}\)
within the full 16-dimensional Bell-state operator space.

In this representation, the corresponding states are  
\begin{equation}
\vec{r}_{\Phi^+}
= \begin{bmatrix}
1\\[4pt]
0
\end{bmatrix}.
\end{equation}
and
\begin{equation}
\vec{r}_{\Psi^-}
= \begin{bmatrix}
0\\[4pt]
1
\end{bmatrix}.
\end{equation}

\begin{figure}[h]
\centering
\hspace{1cm}
\begin{tikzpicture}[
  scale=2.0,                               
  xshift=0.5cm, yshift=-0.5cm,             %
  line cap=round, line join=round, >=Stealth,
  font=\large,                             %
  axis/.style={->, line width=1.0pt},
  rvec/.style={->, line width=1.0pt},
  bvec/.style={->, line width=1.2pt, red},
  mixline/.style={dashed, line width=0.9pt},
  bound/.style={line width=1.1pt, black},
  note/.style={font=\large, inner sep=0.8pt}
]

  \def\xax{1.75}
  \def\yax{1.75}
  \draw[axis] (0,0) -- (\xax,0) 
      node[below right=2pt] {$x_{\scriptscriptstyle|\Phi^+\rangle}$};
  \draw[axis] (0,0) -- (0,\yax)
      node[above left=2pt] {$y_{\scriptscriptstyle|\Psi^-\rangle}$};

  \draw[rvec] (0,0) -- (1,0) node[below=0pt] {$\vec r_{\Phi^+}$};
  \draw[rvec] (0,0) -- (0,1) node[left=0pt] {$\vec r_{\Psi^-}$};

  \draw[mixline] (0,1) -- (1,0);
  \node[note] (mixlbl) at (0.78,1.32) {$\mu\vec r_{\Phi^+} + (1-\mu)\vec r_{\Psi^-}$};
  \draw[->, very thin] (mixlbl.south) -- (0.52,0.52);

  \pgfmathsetmacro{\rtwo}{1.15}
  \pgfmathsetmacro{\kfac}{0.46}
  \draw[bvec] (0,0) -- ({\kfac*2*\rtwo},{-\kfac*2*\rtwo})
      node[pos=0.95, below right=0pt] {$\vec B$};

\pgfmathsetmacro{\off}{1/(2*\rtwo)}
\pgfmathsetmacro{\H}{0.6}
\coordinate (Cplus)  at ({\off/2},{-\off/2});
\coordinate (Cminus) at ({-\off/2},{\off/2});

\draw[bound]
  ($(Cplus)+(-\H,-\H)$) --
  node[pos=0, below left=0.1pt] {$\vec r\!\cdot\!\vec B=2$}   
  ($(Cplus)+(\H,\H)$);

\draw[bound]
  ($(Cminus)+(-\H,-\H)$) --
  node[pos=0, below left=0.1pt] {$\vec r\!\cdot\!\vec B=-2$}
  ($(Cminus)+(\H,\H)$);

\end{tikzpicture}

\caption{\textbf{CHSH value as an operator–state inner product.}
The $x$- and $y$-axes correspond to $|\Phi^+\rangle$ and $|\Psi^-\rangle$.
The unit vectors $\vec r_{\Phi^+}=(1,0)$ and $\vec r_{\Psi^-}=(0,1)$ are shown.
The dashed segment represents the convex mixture $\mu\vec r_{\Phi^+}+(1-\mu)\vec r_{\Psi^-}$, 
whose endpoints correspond to pure states and whose interior points denote mixed states.
The red arrow indicates the operator vector $\vec B$.
The two solid lines are perpendicular to $\vec B$ and mark the classical Bell bounds 
$\vec r\!\cdot\!\vec B=\pm2$; states whose projections lie beyond these bounds exhibit nonlocality.}

\label{fig:bellbounding}

\end{figure}
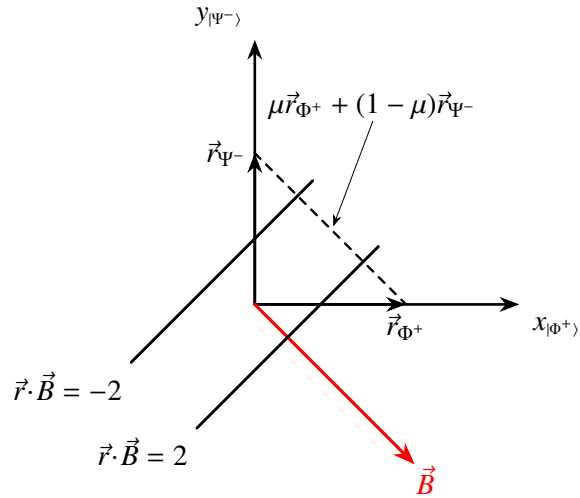

In this subspace, the Gram matrix defined in Sec.~\ref{sec:DataTable} is the identity matrix. Therefore, the expectation value of $\mathcal{B}$ for each state is given by the inner
product between the operator vector and the state vector:
\begin{equation}
\langle \mathcal{B} \rangle_{\Phi^+}
= \vec{B}\!\cdot\!\vec{r}_{\Phi^+}
= 2\sqrt{2},
\qquad
\langle \mathcal{B} \rangle_{\Psi^-}
= \vec{B}\!\cdot\!\vec{r}_{\Psi^-}
= -2\sqrt{2}.
\end{equation}
Hence, $|\Phi^+\rangle$ and $|\Psi^-\rangle$ correspond to the maximal and
minimal eigenstates of the CHSH operator, respectively. The discussion above is illustrated in~\autoref{fig:bellbounding}.




\newpage

\subsection{Box World}
To explore the absolute limits of nonlocality beyond quantum mechanics, we consider the theoretical model known as \textbf{box world}. In this 
framework, any ``box’’ is represented by conditional probabilities subject 
only to the no-signalling condition. This unified perspective allows us to 
compare classical, quantum, and post-quantum correlations on the same 
footing. Within this framework, a fundamental building block is the \textbf{gbit}, which can be viewed as a simple single-system model in box world. Its state space differs from the quantum Bloch sphere. When composed, such systems can give rise to correlations exceeding both classical and quantum limits while remaining fully no-signalling.

Historically, in finite-input and finite-output scenarios, this set of 
correlations has been analysed using tools from convex geometry, giving rise 
to the so-called \textbf{non-signalling polytope}~\cite{Tsirelson1993}.  
In this framework, each allowed correlation corresponds to a point in a 
high-dimensional polytope whose facets encode Bell-type inequalities and 
whose extremal points capture the boundary behaviours permitted by the 
no-signalling principle.  The term ``box world’’ emphasizes the operational 
picture of black boxes defined solely by their input–output behaviour, while 
the geometric viewpoint of the non-signalling polytope highlights the convex 
structure in which these behaviours reside.  In finite scenarios, these two 
perspectives provide closely related descriptions of the same correlation 
space.

Among the nonlocal extremal behaviours compatible with the no-signalling 
constraints, the most celebrated example is the \textbf{Popescu--Rohrlich (PR) box}.  
PR-type boxes attain the algebraic maximum of the CHSH expression (a value of 4) while remaining fully no-signalling.  Although such correlations cannot be realised in nature, their role within box world provides valuable 
insights.



\subsubsection{State space as a non-signalling polytope}
\label{sec:behavior}
Consider two correlated but non-interacting subsystems, each with a finite set of possible measurements and outcomes. For a given state scheme $(M_1, \ldots, M_K; N_1, \ldots, N_L)$ consisting of $K$ measurement settings for Alice and $L$ measurement settings for Bob, let $M_k$ denote the set of possible outcomes of the $k$-th measurement on Alice’s side, and $N_l$ the set of 
possible outcomes of the $l$-th measurement on Bob’s side. We assume that the sets $M_1,\ldots,M_K$ are disjoint and define $M = \bigcup_{k=1}^K M_k$, which collects all possible outcomes of all measurements at Alice's side, and similarly $N = \bigcup_{l=1}^L N_l$. At each side, the marginal probabilities satisfy the normalisation conditions:

\begin{align}
    &\forall k=1,\ldots,K: \quad \sum_{m \in M_k} p_m = 1, \\
    &\forall l=1,\ldots,L: \quad \sum_{n \in N_l} p_n = 1.
\end{align}

Suppose $m \in M_k$ and $n \in N_l$ denote respective outcomes 
for the chosen measurement settings. A \textbf{state} is a family of nonnegative joint probabilities $\{p_{mn}\}_{m \in M_k, n \in N_l}$ satisfying non-signalling conditions:

\begin{align}
    &\forall k,n: \quad \sum_{m \in M_k} p_{mn} = p_n, \label{eq:nonsig_AtoB}  \\
    &\forall l,m: \quad \sum_{n \in N_l} p_{mn} = p_m.\label{eq:nonsig_BtoA}
\end{align}

All possible states together span a convex polytope $X_S$ of dimension:
\begin{equation}
    d = (|M| - K + 1)(|N| - L + 1) - 1,
\end{equation}
where $|M| = |M_1| + \cdots + |M_K|$ and $|N| = |N_1| + \cdots + |N_L|$. The data table shown in~\autoref{tab:operational} may also be referred to as a state. In this case, the general measurements $M$ are interpreted as measurements performed on two parties.

A \textbf{deterministic state} is one in which each $p_{mn}$ is either $0$ or $1$, corresponding to a fixed assignment of outcomes to every possible measurement setting. Every deterministic state is a vertex of the convex polytope $X_\mathrm{S}$. The set of deterministic states is finite and is denoted by $X_\mathrm{DS}$.
Importantly, not every extremal point (vertex) of $X_\mathrm{S}$ is deterministic, that is to say: 
\begin{equation}
X_\mathrm{DS} \subset \mathrm{ex}
(X_\mathrm{S}).
\end{equation}
In general,
\begin{equation}
X_\mathrm{DS} \neq \mathrm{ex}(X_\mathrm{S}), \; \text{or} \; \mathrm{co}(X_\mathrm{DS}) \neq X_\mathrm{S},
\end{equation}
here $X_\mathrm{DS}$ denotes the finite set of deterministic states, while $\mathrm{co}(X_\mathrm{DS})$ represents their convex hull, which is itself a convex subset of the state space. The first inequality states that not all extremal points of $X_S$ are deterministic, and the second emphasizes that taking the convex hull of deterministic states does not recover the full state polytope $X_S$.

In fact, 
\begin{equation}
X_\mathrm{HDS} = \mathrm{co}(X_\mathrm{DS}),  \; \text{or} \; \mathrm{ex}(X_\mathrm{HDS}) = X_\mathrm{DS}.
\end{equation}
Here $X_\mathrm{HDS}$ denotes the set of hidden deterministic states, namely the states admitted within a local hidden-variable theory. The first identity expresses that any hidden deterministic state can be written as a convex combination of deterministic states, while the second emphasizes that the deterministic states are precisely the extremal points of $X_\mathrm{HDS}$.

In general,
\begin{equation}
X_\mathrm{HDS} \neq X_\mathrm{S}
\end{equation}
This separation allows one to formulate Bell-type inequalities that distinguish local hidden-variable states from more general no-signalling states.

\subsubsection{Example: The simplest scenario $(2+2) \times (2+2)$}

We illustrate the state space and the classical Bell-type inequalities with the
simplest nontrivial scenario: two inputs per party and two binary outcomes.
All conditional probabilities can be arranged in a $4\times4$ array
\[
P = \left(
\begin{array}{cc|cc}
p_{11} & p_{12} & p_{13} & p_{14} \\
p_{21} & p_{22} & p_{23} & p_{24} \\
\hline
p_{31} & p_{32} & p_{33} & p_{34} \\
p_{41} & p_{42} & p_{43} & p_{44}
\end{array}
\right),
\]
with $p_{m1}+p_{m2}=p_{m3}+p_{m4}$, $p_{1n}+p_{2n}=p_{3n}+p_{4n}.$
The scheme admits rich symmetries (8 ``horizontal'' and 8 ``vertical''), 
yielding a symmetry group of size $64$. 
For instance, considering horizontal permutations of the probability table, 
we obtain the following subgroup of $8$ elements:
\[
\begin{aligned}
&(p_1\,p_2 \mid p_3\,p_4), \quad 
(p_1\,p_2 \mid p_4\,p_3), \quad 
(p_2\,p_1 \mid p_3\,p_4), \quad 
(p_2\,p_1 \mid p_4\,p_3), \\[6pt]
&(p_3\,p_4 \mid p_1\,p_2), \quad 
(p_3\,p_4 \mid p_2\,p_1), \quad 
(p_4\,p_3 \mid p_1\,p_2), \quad 
(p_4\,p_3 \mid p_2\,p_1).
\end{aligned}
\]
These operations differ only by permutations of indices.
Both the state polytope $X_S$ and $X_{\mathrm{HDS}}$ are $8$-dimensional convex polytopes (can be calculated from the probability table: $d=(4-2+1)\times(4-2+1)-1=8$);

\paragraph{\textbf{Extremal states.}}\label{para:extremal}
Vertices of $X_{\mathrm{HDS}}$ coincide with the deterministic states (which are also vertices of $X_S$). 
A representative deterministic vertex is
\[
\left(
\begin{array}{cc|cc}
1&0&1&0\\
0&0&0&0\\
\hline
1&0&1&0\\
0&0&0&0
\end{array}
\right),
\] while the others can be obtained by symmetry. Equivalently, a deterministic vertex corresponds to the case where each party outputs a fixed value for each input, yielding $2^2 \times 2^2 = 16$ such vertices in total.

In addition, $X_S$ contains $8$ non-deterministic extremal states. 
A representative one is
\[
\frac{1}{2}
\left(
\begin{array}{cc|cc}
1&0&1&0\\
0&1&0&1\\
\hline
1&0&0&1\\
0&1&1&0
\end{array}
\right),
\] which corresponds to a Popescu--Rohrlich (PR) box, to be discussed in detail later.
The others are obtained by symmetry.
Therefore, $X_{\mathrm{HDS}}$ has $16$ vertices, while $X_S$ has $16+8=24$ vertices.

\subsubsection{The dual cone of the non-signalling polytopes}

A state $x$, specified by the probabilities $p_{mn}$, can be evaluated by a linear functional
\begin{equation}
f(x) = \sum_{m,n} \lambda_{mn} p_{mn}
      = \sum_{k,l} \sum_{m \in M_k, n \in N_l} f_{kl}(m,n)p_{mn},
\end{equation}
with real coefficients $\lambda_{mn}$ or equivalently functions $f_{kl}$.  

For a deterministic state $(\alpha_1,\dots,\alpha_K;\beta_1,\dots,\beta_L)$, this reduces to
\begin{equation}
f(\alpha_1,\dots,\alpha_K;\beta_1,\dots,\beta_L) = \sum_{k,l} f_{kl}(\alpha_k,\beta_l).
\end{equation}

The collection of all such $f$ forms a $(d+1)$-dimensional vector space (the extra dimension comes from constant functions).  

Those functions $f$ that are nonnegative on the local hidden-variable polytope $X_\mathrm{HDS}$ constitute a convex polyhedral cone. Being the cone dual to $X_\mathrm{HDS}$, it is denoted by $X_\mathrm{HDS}^\circ$, i.e.
\begin{equation}
f \in X_\mathrm{HDS}^\circ \;\;\Longleftrightarrow\;\; f(x)\geq 0 \quad \forall x \in X_\mathrm{HDS}.
\end{equation}

Conversely,
\begin{equation}
x \in X_\mathrm{HDS} \;\;\Longleftrightarrow\;\; f(x)\geq 0 \quad \forall f \in X_\mathrm{HDS}^\circ.
\end{equation}

Since $X_S \supset X_\mathrm{HDS}$ and $X_S \neq X_\mathrm{HDS}$, their dual cones satisfy
\begin{equation}
X_S^\circ \subset X_\mathrm{HDS}^\circ, \quad X_S^\circ \neq X_\mathrm{HDS}^\circ.
\end{equation}


The reason is that the dual cone is defined as the set of all linear functionals that remain non-negative over the corresponding state set. Since $X_S \supset X_\mathrm{HDS}$, any functional $f$ in $X_S^\circ$ must be non-negative on a larger set of states than those in $X_\mathrm{HDS}^\circ$. This imposes stronger constraints on $f$, thereby producing a smaller admissible cone.

Any $f \in X_\mathrm{HDS}^\circ$ and not in $X_S^\circ$ corresponds to a \emph{classical Bell-type inequality}
\begin{equation}
f(x) \geq 0,
\end{equation}
which all local hidden-variable states obey but which may be violated by more general no-signalling states. Here, the term ``Bell-type inequality'' is used in the generalised geometric sense: it refers to any valid linear inequality for the local polytope, not only to the original Bell inequality or a specific CHSH-type form.

In finite case, a classical Bell-type inequality is called \emph{extremal} if $f$ lies on an extremal ray of the cone, i.e.
\begin{equation}
f \in \mathrm{ext}(X_\mathrm{HDS}^\circ) \setminus X_S^\circ.
\end{equation}

Considering all extremal classical Bell-type inequalities 
$f_1(x)\geq 0,\dots,f_\nu(x)\geq 0$, we obtain
\begin{equation}
\forall x \in X_S, \quad
x \in X_\mathrm{HDS}
\;\;\Longleftrightarrow\;\;
f_1(x)\geq 0,\dots,f_\nu(x)\geq 0.
\end{equation}

Thus, these inequalities form a full and non-redundant set of consequences of local realism for the given state scenario.

\paragraph{\textbf{Example: The (2+2)×(2+2) scheme in the dual picture}}
The dual space can be identified with the $9$-dimensional $(d+1)$ family of affine-linear functionals 
on four binary variables $\alpha_1,\alpha_2,\beta_1,\beta_2\in\{\pm1\}$:
\begin{equation}
\label{eq:f-affine}
f(\alpha,\beta)=
a_1\alpha_1+a_2\alpha_2+b_1\beta_1+b_2\beta_2
+c_{11}\alpha_1\beta_1+c_{12}\alpha_1\beta_2+c_{21}\alpha_2\beta_1+c_{22}\alpha_2\beta_2+d,
\end{equation}
where the coefficients are linear combinations of the probabilities. Equivalently,
\begin{equation}
f(\alpha,\beta,\gamma)=a_1 \alpha_1 + a_2 \alpha_2 
+ b_1 \beta_1 + b_2 \beta_2 
+ c_{11} \gamma_{11} + c_{12} \gamma_{12} 
+ c_{21} \gamma_{21} + c_{22} \gamma_{22} 
+ d
\label{eq:linear_function}
\end{equation}
A convenient choice of coordinates is
\begin{equation}
\label{eq:alphagamma}
\alpha_1=(p_{11}+p_{12})-(p_{21}+p_{22})=(p_{13}+p_{14})-(p_{23}+p_{24}),\qquad
\gamma_{11}=p_{11}-p_{12}-p_{21}+p_{22},
\end{equation}
and analogues by symmetry for $\alpha_2,\beta_1,\beta_2,\gamma_{12},\gamma_{21},\gamma_{22}$.
With these, faces of the polytopes appear as linear Bell-type inequalities. For instance,
\begin{equation}
\label{eq:faceXB}
\alpha_1+\beta_1-\gamma_{11}\;\le 1
\quad\text{(a face of $X_S$),}
\end{equation}
and, by symmetry, similar ones such as $\alpha_1+\beta_1+\gamma_{11}\le1\ldots$. These faces are also faces for $X_{HDS}$, while $X_{HDS}$ possesses additional ones of the form
\begin{equation}
\label{eq:CHSH-face}
\gamma_{11}+\gamma_{12}+\gamma_{21}-\gamma_{22}\;\le 2
\quad\text{(a face of $X_{\mathrm{HDS}}$, i.e.\ an extremal classical Bell-type inequality).}
\end{equation} and by symmetry, there will be 8 extra faces. Therefore, $X_S$ has $16$ faces and $X_{HDS}$ has $24$ faces.


\subsubsection{Relation to quantum state space}
In the previous section we described states $p_{mn}$ in a purely operational way, treating them as conditional probability distributions subject only to normalisation and no-signalling constraints. We now impose the further restriction that the system is quantum mechanical.

Concretely, a state admits a \emph{quantum realization} if there exists a density 
matrix $W$ on a bipartite Hilbert space $\mathcal{H}_1\otimes\mathcal{H}_2$ together 
with positive operators $\{F_m\}$ and $\{F_n\}$ forming POVMs on each subsystem, such that
\begin{equation}
p_{mn}=\mathrm{Tr}(F_m F_n W), \qquad 
W\ge 0,\ \mathrm{Tr}(W)=1,
\label{eq:WCondition}
\end{equation}
and
\begin{equation}
\forall m\in M\quad F_M\geq 0,\quad \forall n\in N\quad F_M\geq 0,\quad
\sum_{m\in M_k} F_m = 1, \quad 
\sum_{n\in N_l} F_n = 1, \quad 
F_m F_n = F_n F_m.
\label{eq:FCondition}
\end{equation}
These are the minimal quantum requirements. 

In the maximal requirement (projective) case, one can write
\begin{equation}
p_{mn} = \langle \Psi | P_m\otimes P_n |\Psi\rangle,
\end{equation}
where $|\Psi\rangle\in\mathcal{H}_1\otimes\mathcal{H}_2$ is an entangled state vector and $\langle\Psi|\Psi\rangle=1$. The operators $\{P_m\}$ are projectors satisfying



\begin{equation}
\begin{aligned}
P_m &: \mathcal{H}_1 \rightarrow \mathcal{H}_1, \quad m \in M, \\
P_m^2 &= P_m = P_m^*, \\
\sum_{m \in M_k} P_m &= \mathbb{I}, \quad \forall k=1,\ldots,K, \\
P_{m_1} P_{m_2} &= 0, \quad m_1 \neq m_2,\;\; m_1,m_2 \in M_k.
\end{aligned}
\end{equation}

The operators $\{P_n\}$ satisfy the same conditions. Thus, for each setting, $\{P_m\}$ and $\{P_n\}$ define projective measurements. From these projectors, one may form Hermitian observables
\begin{equation}
A_k = \sum_{m\in M_k} m P_m,
\qquad 
B_l = \sum_{n\in N_l} n P_n,
\end{equation}
where $m$ and $n$ are real numbers of the measurement outcomes. 
In this way, the probabilities $p_{mn}$ are just the joint distribution of the two observables.
In general, these observables need not commute,
\begin{equation}
A_{k_1} A_{k_2} \neq A_{k_2} A_{k_1}, 
\qquad 
B_{l_1} B_{l_2} \neq B_{l_2} B_{l_1}.
\end{equation}

The class of states generated by the minimal quantum formulation coincides with that of the maximal (projective) formulation, despite the former being formally more general, including (i) mixed states, (ii) general POVMs, (iii) correlations with unobserved subsystems, and (iv) non-tensor-product operator algebras.

A state that admits a representation of the form ~\eqref{eq:WCondition} or equivalently ~\eqref{eq:FCondition}, is called a \emph{quantum state}. The set of all quantum states for a fixed scenario 
$(M_1,\ldots,M_K; N_1,\ldots,N_L)$ is a convex compact body, denoted $X_{QS}$, satisfying
\begin{equation}
X_{\mathrm{HDS}} \subset X_{QS} \subset X_S,
\end{equation} 
with $X_{\mathrm{HDS}} \neq X_{QS} \neq X_S$ in general. The separation between $X_{\mathrm{HDS}} \neq X_{QS}$ underlies the existence of classical Bell-type inequalities, while $X_{QS} \neq X_S$ is equivalent to the existence of quantum Bell-type inequalities. Classical Bell-type inequalities are defined under the assumption that all observables commute, in contrast to quantum Bell-type inequalities, which only require commutativity between observables associated with different subsystems.

The above discussion suggests several properties of the quantum state set $X_{QS}$. The non-signalling state set $X_S$ and the local hidden deterministic set $X_{\mathrm{HDS}}$ are \emph{polytopes}. 
In contrast, the quantum set $X_{QS}$ is a \emph{convex compact body}, which in general is not a polytope. 
Moreover, $X_{QS}$ contains extremal quantum states that cannot be expressed as probabilistic mixtures of other quantum states. 
Since $X_{QS}$ is not a polytope, it cannot be described by finitely many linear inequalities.

For any linear functional of the form
\begin{equation}
  f(x) = \sum_{k,l}\sum_{m\in M_k}\sum_{n\in N_l} f_{kl}(m,n)\,p_{mn},
\end{equation}
the quantum maximum is given by
\begin{equation}
  \max_{x\in X_{QS}} f(x)
  \;=\;
  \max \Bigg(
    \max\,\text{spec} \sum_{k,l} f_{kl}(A_k,B_l)
  \Bigg),
\label{eq:spec}
\end{equation}
The left-hand side represents the quantum bound of $f$. 
On the right-hand side, $f_{k,l}(A_k,B_l)$ denotes the operator in $\mathcal{H}_1 \otimes \mathcal{H}_2$ obtained by applying the scalar function $f_{k,l}$ to the pair of commuting operators $A_k,B_l$ introduced above. 
The inner maximization, written as ``$\max \, \text{spec}$,'' corresponds to taking the largest eigenvalue of the operator $\sum_{k,l} f_{kl}(A_k,B_l)$. 
The outer maximization is then taken over all admissible choices of operator collections $(A_1,\ldots,A_K; \, B_1,\ldots,B_L)$ acting on $\mathcal{H}_1,\mathcal{H}_2$, respectively, with the condition that the eigenvalues of each $A_k$ lie in $M_k$ and those of each $B_l$ lie in $N_l$.

By compactness of $X_{QS}$, the image of any linear functional $f$ is bounded and closed, hence the quantum bound always exists.

\vspace{6pt}

\subsubsection{The Gbit}
In standard quantum mechanics, the state of a qubit is characterized by its Bloch vector, whose components correspond to the expectation values of the Pauli operators $(X,Y,Z)$. These three spin measurements can be regarded as a set of fiducial measurements: a set of reference measurements whose outcome probabilities are sufficient to determine the state of the system. Moreover, in quantum theory, the values of these components are constrained by quantum positivity: it is not possible for all three Pauli expectation values to simultaneously take the values $\pm 1$. This is a fundamental restriction on the state space, and it geometrically manifests as the Bloch sphere, where valid states lie within a unit ball.

But what if we relaxed this constraint?

Suppose we allow all three Pauli expectation values to simultaneously take the values $\pm1$. The resulting state space would no longer be the Bloch ball, but a cube with eight extremal points, one at each corner. This defines an alternative state space to that of a qubit, called a \textbf{gbit}, where ``g'' stands for generalised. The gbit therefore offers a first glimpse of state spaces beyond the Bloch ball and points toward the broader framework of generalised probabilistic theories (GPTs).

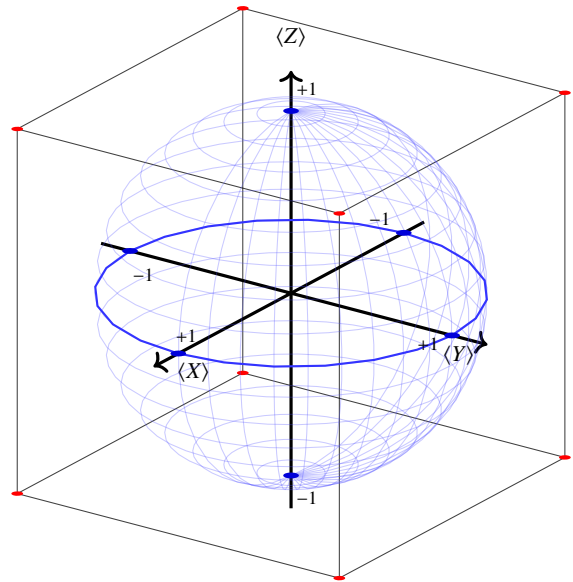
\begin{figure}[htbp]
\centering

\tdplotsetmaincoords{68}{125}

\begin{tikzpicture}
\begin{scope}[tdplot_main_coords, scale=2.6]

\draw[very thick] (-1.18,0,0) -- (0,0,0);
\draw[->, very thick] (0,0,0) -- (1.22,0,0);

\draw[very thick] (0,-1.18,0) -- (0,0,0);
\draw[->, very thick] (0,0,0) -- (0,1.22,0);

\draw[very thick] (0,0,-1.18) -- (0,0,0);
\draw[->, very thick] (0,0,0) -- (0,0,1.22);

\node[anchor=west, xshift=6pt, yshift=-2pt] at (1.22,0,0) {$\langle X\rangle$};
\node[anchor=east, xshift=-2pt, yshift=-4pt] at (0,1.22,0) {$\langle Y\rangle$};
\node[anchor=south, yshift=6pt] at (0,0,1.22) {$\langle Z\rangle$};

\fill (0,0,0) circle (0.025);

\coordinate (A) at (-1,-1,-1);
\coordinate (B) at ( 1,-1,-1);
\coordinate (C) at ( 1, 1,-1);
\coordinate (D) at (-1, 1,-1);

\coordinate (E) at (-1,-1, 1);
\coordinate (F) at ( 1,-1, 1);
\coordinate (G) at ( 1, 1, 1);
\coordinate (H) at (-1, 1, 1);

\draw[black, opacity=0.75] (A)--(B)--(C)--(D)--cycle;
\draw[black, opacity=0.75] (E)--(F)--(G)--(H)--cycle;
\draw[black, opacity=0.75] (A)--(E);
\draw[black, opacity=0.75] (B)--(F);
\draw[black, opacity=0.75] (C)--(G);
\draw[black, opacity=0.75] (D)--(H);

\foreach \p in {A,B,C,D,E,F,G,H}{
    \fill[red] (\p) circle (0.03);
}

\foreach \theta in {0,10,...,170}{
    \draw[blue!70, opacity=0.30]
    plot[variable=\phi, domain=0:360, samples=40]
    (
        {sin(\theta)*cos(\phi)},
        {sin(\theta)*sin(\phi)},
        {cos(\theta)}
    );
}

\foreach \phi in {0,10,...,170}{
    \draw[blue!70, opacity=0.30]
    plot[variable=\theta, domain=0:180, samples=20]
    (
        {sin(\theta)*cos(\phi)},
        {sin(\theta)*sin(\phi)},
        {cos(\theta)}
    );
}

\draw[blue!80, thick]
plot[variable=\t, domain=0:360]
({cos(\t)},{sin(\t)},0);

\foreach \p in {
    (1,0,0),(-1,0,0),
    (0,1,0),(0,-1,0),
    (0,0,1),(0,0,-1)
}{
    \fill[blue!85!black] \p circle (0.04);
}

\node[font=\scriptsize, anchor=north west, xshift=-4pt, yshift=12pt] at (1,0,0) {$+1$};
\node[font=\scriptsize, anchor=south east, xshift=-2pt] at (-1,0,0) {$-1$};

\node[font=\scriptsize, anchor=east, xshift=-2pt,yshift=-3pt] at (0,1,0) {$+1$};
\node[font=\scriptsize, anchor=west, xshift=-2pt,yshift=-9pt] at (0,-1,0) {$-1$};

\node[font=\scriptsize, anchor=south, xshift=6pt, yshift=3pt] at (0,0,1) {$+1$};
\node[font=\scriptsize, anchor=north, xshift=6pt, yshift=-3pt] at (0,0,-1) {$-1$};

\end{scope}
\end{tikzpicture}

\caption{\textbf{Gbit state space with embedded Bloch sphere.} The cube represents the extremal deterministic state space, while the Bloch ball represents the quantum state space constrained by positivity of expectation values. The Bloch sphere is inscribed in the cube and touches the six faces at their centre points.}
\label{fig:gbit_bloch}
\end{figure}

In the GPT framework, a system specified by $n$ fiducial measurements with $k$ outcomes each is often referred to as an $(n,k)$-system~\cite{JBINFOPRO}. Thus, both the qubit and the gbit constitute $(3,2)$-systems, each defined by three binary fiducial measurements, as illustrated in~\autoref{fig:gbit_bloch}. In this sense, the gbit can be viewed as a simple extremal structure within the non-signalling polytope framework. The difference between the gbit and the qubit is that quantum positivity restricts the qubit expectation values to the Bloch ball, whereas the gbit permits all deterministic assignments of the three fiducial outcomes.

\textbf{Transformations.}  
In GPTs, a system's state space has a specific geometric structure—for example, a sphere for a qubit, and a cube for a gbit. An allowed transformation within this framework is any operation that maps the state space onto itself, preserving its geometric shape and convex structure.

For gbits, whose state space is a cube, the allowed reversible transformations are strictly limited to the discrete symmetries of the cube. Concretely, these are operations such as swapping measurement axes or reversing the direction of an axis (i.e., changing the sign of measurement outcomes). Such discrete transformations form a finite group known as the symmetry group of the cube.

This discrete set of allowed transformations contrasts sharply with quantum mechanics, where transformations form continuous groups allowing smooth transitions between states. Thus, gbits differ fundamentally from qubits in their dynamical structure.


\textbf{Composition.}
Composing multiple gbits within the GPT framework is not merely a matter of stacking independent single-system state spaces; the structure of the joint state space crucially depends on the physical constraints imposed on the composite system. A key constraint is the no-signalling principle, which ensures that measurement outcomes on one subsystem remain independent of measurement choices made on distant subsystems. Remarkably, this principle alone is permissive enough to allow so-called super-quantum correlations—joint behaviors that go beyond what is possible in quantum theory. 

The PR-box is a canonical example of a bipartite state of two gbits; it achieves the algebraic maximum violation (equal to 4) of the CHSH inequality—well beyond the quantum Tsirelson bound of $2\sqrt{2}$. Investigating these extreme correlations helps elucidate fundamental distinctions between quantum and broader post-quantum theoretical landscapes.

\subsubsection{The PR Box}

Quantum mechanics exhibits nonlocal correlations, yet these correlations are not maximal under the no-signalling constraint. This raises a natural question: what correlations are possible in theories that respect only the no-signalling principle—that information cannot be transmitted faster than light?

Addressing this question, Popescu and Rohrlich proposed a theoretical model~\cite{PRBOX}, now known as the Popescu--Rohrlich (PR) box, which achieves correlations exceeding those allowed by quantum mechanics while still respecting no-signalling.

This operational principle defines a landscape larger than quantum mechanics, since no-signalling alone does not constrain correlations to the quantum limit. The PR box is a canonical example of such post-quantum correlations, illustrating that the no-signalling principle by itself permits correlations that are strictly stronger than quantum correlations.

To formally capture the strength of nonlocal correlations, a commonly used tool is the CHSH inequality, a variant of Bell’s inequality. In this framework, two parties each choose between two binary measurements, and compare their outcomes. For any local realistic theory, the following combination of correlation functions is bounded:
\begin{equation}
    |E(x,y) + E(x,y') + E(x',y) - E(x',y')| \le 2
\end{equation}

Here, $E(x, y)$ denotes the correlation function for inputs $x$ and $y$, defined as:
\begin{equation}
    E(x, y) = \sum_{a, b} (-1)^{a \oplus b} \cdot P(a, b \mid x, y),
\end{equation}
where $P(a, b \mid x, y)$ is the probability of obtaining outputs $a$ and $b$ given measurement inputs $x$ and $y$, and $(-1)^{a \oplus b}$ takes the value $+1$ when $a = b$ and $-1$ when $a \ne b$.

In contrast, quantum mechanics allows this quantity to reach up to Tsirelson’s bound $2\sqrt{2}$, but no more. This limitation reflects the fundamental structure of quantum theory. However, post-quantum models can reach the algebraic maximum of 4, indicating a deeper level of nonlocality.

A prominent example is the Popescu--Rohrlich (PR) box—an abstract device constructed from two gbits, analogous to how an EPR pair is composed of two entangled qubits. The PR box exemplifies the kind of “super-quantum” correlations that respect no-signalling yet reach the algebraic maximum violation of Bell inequalities.

One can define the Popescu--Rohrlich (PR) box in an operationally simple way. Two observers, Alice and Bob, each receive a binary input $x, y \in \{0,1\}$ and return binary outputs $a, b \in \{0,1\}$, constrained by the relation:
\begin{equation}
    a \oplus b = x \cdot y
\end{equation}
This condition ensures perfect correlation or anti-correlation between the outputs depending on the inputs. Notably, the correlations generated by this rule achieve the algebraic maximum value 4 in the CHSH expression, while still respecting the no-signalling condition.

Now we compute each correlator $E(x,y)$ for the four input settings used in the CHSH expression:
\begin{equation}
\text{CHSH} = |E(0,0) + E(0,1) + E(1,0) - E(1,1)|
\end{equation}

When $(x,y) = (0,0)$, we have $x \cdot y = 0$, so $a \oplus b = 0$, which implies $a=b$. Therefore, the valid outcomes are $(0,0)$ and $(1,1)$, each occurring with probability $\frac{1}{2}$. For both outcomes, $(-1)^{a \oplus b}=1$, and hence
\[
E(0,0)=1.
\]

When $(x,y) = (0,1)$, we again have $x \cdot y = 0$, so $a \oplus b = 0$ and thus $a=b$. By the same reasoning as above, it follows that
\[
E(0,1)=1.
\]

When $(x,y) = (1,0)$, we still have $x \cdot y = 0$, so $a \oplus b = 0$ and hence $a=b$. Therefore,
\[
E(1,0)=1.
\]

Finally, when $(x,y) = (1,1)$, we have $x \cdot y = 1$, so $a \oplus b = 1$, which implies $a \ne b$. The valid outcomes are then $(0,1)$ and $(1,0)$, each with probability $\frac{1}{2}$. For both cases, $(-1)^{a \oplus b}=(-1)^1=-1$, and thus
\[
E(1,1)=-1.
\]

\noindent Putting it all together:
\begin{equation}
\text{CHSH} = |1 + 1 + 1 - (-1)| = 4
\end{equation}

The Popescu--Rohrlich (PR) box is closely connected to the structure of the non-signalling polytope. As discussed in Sec.~\ref{sec:behavior}, the set of no-signalling correlations forms a convex polytope whose extremal points correspond to fundamental behaviors. Within this geometric framework, PR boxes are precisely the non-deterministic extremal points of the non-signalling polytope. One representative example used in Sec.~\ref{para:extremal} is
\[
\frac{1}{2}
\left(
\begin{array}{cc|cc}
1 & 0 & 1 & 0\\
0 & 1 & 0 & 1\\
\hline
1 & 0 & 0 & 1\\
0 & 1 & 1 & 0
\end{array}
\right),
\]
where each block corresponds to a pair of measurement settings, which can be identified with the inputs $(x,y)$ introduced previously. The remaining ones can be obtained by symmetry, giving a total of eight non-deterministic extremal points. This observation shows that PR boxes are not ad hoc constructions, but instead emerge naturally from the extremal structure of the theory.

If one considers the marginal statistics of the PR box associated with a single party, the resulting local distributions correspond exactly to a gbit structure. For consistency with the preceding notation, we identify the measurement settings $x,y\in\{0,1\}$ with the fiducial observables $X$ and $Z$, respectively, and the binary outcomes $a,b\in\{0,1\}$ with the corresponding measurement results.

More explicitly, the conditional structure induced by Alice’s measurements can be summarised as follows. When Alice performs measurement $x=0$ (i.e., $X$) and obtains outcome $a=1$, Bob’s conditional expectation values satisfy $\langle X\rangle_B = 1$ and $\langle Z\rangle_B = 1$. When Alice measures $x=0$ and obtains $a=0$, the corresponding conditional state is $\langle X\rangle_B = 0$ and $\langle Z\rangle_B = 0$.

Similarly, when Alice measures $x=1$ (i.e., $Z$) and obtains outcome $a=1$, Bob’s conditional expectations are $\langle X\rangle_B = 1$ and $\langle Z\rangle_B = 0$, whereas for $a=0$ one obtains $\langle X\rangle_B = 0$ and $\langle Z\rangle_B = 1$.

These four conditional states form the vertices of a square in the $(\langle X\rangle_B,\langle Z\rangle_B)$ plane, thereby realising a two-dimensional gbit state space.

\subsubsection{Dynamics in box world}

Having introduced the state space of box world, we now turn to the structure of allowed dynamics. In the GPT framework, transformations are represented by linear maps that preserve the set of allowed states. That is, 
any physically admissible transformation $M$ must satisfy
\begin{equation}
M(S) \subseteq S,
\end{equation}
where $S$ denotes the state space.

The setting considered here corresponds to what is known in the literature as 
generalised non-signalling theory (GNST), where the state space is precisely the 
non-signalling polytope. The constraint above therefore requires that any transformation 
maps this polytope into itself.

\textbf{Single-system dynamics.}

For an $(n,k)$ system, the normalised state space forms a convex polytope whose extremal 
points correspond to deterministic assignments of outcomes to fiducial measurements. 
Any allowed transformation must map this polytope into itself.

This requirement is highly restrictive. In fact, all normalisation-preserving transformations 
reduce to relabellings of fiducial measurements and outcomes, together with convex combinations 
of such operations. Equivalently, any transformation can be written in block form
\begin{equation}
M =
\left(
\begin{array}{c|c|c}
M_{11} & \cdots & M_{1n} \\ \hline
\vdots & \ddots & \vdots \\ \hline
M_{n1} & \cdots & M_{nn}
\end{array}
\right),
\end{equation}
where each block satisfies
\begin{equation}
M_{ij} = \alpha_{ij} S_{ij}, \qquad 0 \le \alpha_{ij} \le 1, \quad \sum_j \alpha_{ij} = 1,
\end{equation}
with $S_{ij}$ stochastic matrices.

Thus, despite the highly nonclassical structure of the state space, the dynamics of single systems 
is essentially classical: it consists only of permutations, relabellings, and probabilistic mixtures.

A similar restriction applies to measurements, which reduce to fiducial measurements and their 
relabelled or mixed versions. No genuinely new measurement structures arise.

\textbf{Composite systems dynamics.}

For composite systems, the structure of admissible dynamics remains highly restricted, but its form is more subtle than in the single-system case. In particular, for bipartite systems consisting of two gbits in GNST, Barrett showed that any normalisation-preserving transformation can be characterized operationally through the joint statistics obtained after the transformation~\cite{JBINFOPRO}. More precisely, if the transformed system is followed by fiducial measurements $X$ and $Y$ on the two subsystems, with final outcomes $a$ and $b$, then the joint distribution $P(a,b\mid X,Y)$ can always be written as a \textbf{convex combination} of protocols of the following kind: first perform a fiducial measurement $X'$ on one subsystem, where $X'$ may depend on $X$ and $Y$, and denote its outcome by $a'$; then perform a fiducial measurement $Y'$ on the other subsystem, where $Y'$ may depend on $X$, $Y$, and $a'$, and denote its outcome by $b'$; finally, the output pair $(a,b)$ is a function of $X$, $Y$, $a'$, and $b'$. 

A closely related restriction holds for measurements on bipartite systems of two gbits. 
In this case, any such measurement is a \textbf{convex combination} of protocols in which 
one first performs a fiducial measurement $X$ on one subsystem, obtaining an outcome $a'$, 
and then performs a fiducial measurement $Y$ on the other subsystem, where $Y$ may depend 
on $a'$, obtaining an outcome $b'$. The final measurement outcome is then a function of 
$a'$ and $b'$.

Here, the dependence of the second measurement only on $a'$ reflects the fact that the measurement procedure itself is fixed, so that any dependence on external measurement choices can be absorbed into the definition of the protocol. Consequently, the only nontrivial information available for conditioning the second measurement is the outcome 
$a'$ obtained in the first step.

These results show that, even for bipartite systems admitting maximally nonlocal states such as PR-box correlations, the allowed transformations and measurements remain decomposable into sequential fiducial procedures with classical post-processing. Thus, the nonlocal richness of the state space is not accompanied by an equally rich class of joint dynamical operations. This is consistent with later results showing that reversible dynamics in box world are essentially trivial~\cite{gross2010all}, consisting only of local relabellings and permutations of subsystems. In particular, reversible transformations map pure product states to pure product states, and hence PR-box correlations cannot be obtained from product states through reversible dynamics.

This restriction can also be understood as a consequence of the large state space allowed in GNST. As noted by Barrett~\cite{JBINFOPRO}, admissible transformations must map every allowed state to another allowed state. In this sense, allowing a larger state space can impose stronger constraints on possible transformations. In box world, where the full non-signalling polytope is allowed, this requirement leaves only a limited class of dynamics, namely sequential fiducial procedures with classical conditioning and post-processing.

This also clarifies the contrast with quantum theory. Quantum theory allows a smaller set of states than the full non-signalling polytope, but it admits a richer class of continuous reversible transformations, such as unitary evolutions. Thus, the comparison suggests that the operational power of a theory is not determined only by how nonclassical its state space is, but also by what transformations and measurements are allowed.






\newpage

\section{Hamiltonian mechanics and phase-space in GPTs}

Defining Hamiltonians in GPTs requires generalising their dual role in classical and quantum mechanics: as observables representing energy and as generators of reversible dynamics. Recent work has introduced a set of desiderata—observability, generation of dynamics, invariance under time evolution, and recovery of standard quantum behavior—that constrain how Hamiltonians can be defined within GPTs. This framework not only ensures that energy retains its physical meaning beyond classical and quantum mechanics but also provides a systematic route for constructing generators of dynamics in non-Hilbert-space theories.

A unifying description of Hamiltonian mechanics across classical, quantum, and post-quantum theories can be achieved with generalised phase-space distributions. In classical mechanics, time evolution is governed by the Poisson bracket with the Hamiltonian, while in quantum mechanics it is expressed through the Moyal bracket acting on the Wigner function. More generally, post-quantum theories allow representations with kernels extending these brackets. In all cases, the Hamiltonian plays the central role in encoding energy and driving reversible evolution. By adopting generalised quasiprobability distributions, GPTs capture these structures in a unified framework.

Wigner function negativity highlights unique features of quantum theory. Unlike classical distributions, the Wigner distribution can assume negative values, enabling behaviors impossible in classical statistics. Such negativity manifests in operationally accessible scenarios, serving as a signature of nonclassicality beyond Hilbert spaces. The deep link between negativity and contextuality reveals structural distinctions among classical, quantum, and post-quantum phase-space theories.

Through these topics, this section illustrates how GPTs provide a principled extension of Hamiltonian mechanics and phase space theory, enabling us to reexamine foundational issues in dynamics and thermodynamics. The tools introduced here aim to prepare the reader for engaging with current research at the intersection of quantum foundations and information theory.





\subsection{Operational Characterization of Hamiltonians in GPTs}

To extend the concept of Hamiltonians beyond classical and quantum mechanics, we require a definition that is meaningful within the GPTs framework. The Hamiltonian should retain its dual role as both an observable and a generator of time evolution. Following previous works~\cite{barnum2014higher, 018GenHam}, we introduce four desiderata that any such definition should satisfy:
\begin{itemize}
    \item \textbf{(OBS)}: The Hamiltonian must be an observable---that is, a linear functional on the state space expressible as a linear combination of measurement effects.
    \item \textbf{(GEN)}: The Hamiltonian must (at least partially) determine the generator of time evolution, ensuring that dynamics arise from a one-parameter group of reversible transformations.
    \item \textbf{(INV)}: The expectation value of the Hamiltonian must be conserved under time evolution, generalising energy conservation.
    \item \textbf{(QUAN)}: In quantum GPTs, the definition must reproduce standard quantum dynamics (e.g., unitary evolution generated by the Hermitian Hamiltonian).
\end{itemize}

These desiderata each impose nontrivial structural constraints. \textbf{OBS} requires that the Hamiltonian be linearly decomposable in terms of effects. \textbf{GEN} implies that the generator of dynamics---represented as a real antisymmetric matrix---must be determined (at least partially) by the Hamiltonian. \textbf{INV} further constrains this relationship: in theories where states are real vectors \( \vec{\rho} \), conservation of \( \vec{H} \cdot \vec{\rho} \) under time evolution imposes orthogonality between the generator and the Hamiltonian vector. Finally, \textbf{QUAN} demands that this framework reduce to the usual Hamiltonian-generated dynamics in the case of quantum state spaces.

To illustrate the construction, consider GPTs with three-dimensional state spaces, where normalised states are real vectors \( \vec{\rho} \in \mathbb{R}^3 \). The space of antisymmetric generators is spanned by the standard \( \mathfrak{so}(3) \) basis:
\[
L_x = \begin{pmatrix}
0 & 0 & 0 \\
0 & 0 & -1 \\
0 & 1 & 0
\end{pmatrix}, \quad
L_y = \begin{pmatrix}
0 & 0 & 1 \\
0 & 0 & 0 \\
-1 & 0 & 0
\end{pmatrix}, \quad
L_z = \begin{pmatrix}
0 & -1 & 0 \\
1 & 0 & 0 \\
0 & 0 & 0
\end{pmatrix}.
\]
A Hamiltonian is then identified with a real vector \( \vec{H} = (H_1, H_2, H_3) \), and the generator of time evolution is constructed as
\[
A = H_1 L_x + H_2 L_y + H_3 L_z.
\]

\begin{figure}[h!]
\centering
\tdplotsetmaincoords{70}{120}
\begin{tikzpicture}[
  tdplot_main_coords, 
  scale=2.5,
  line cap=round,
  line join=round,
  >=Stealth,
  font=\normalsize
]

  \draw[thick, tdplot_screen_coords] (0,0) circle (0.8);

  \draw[->, line width=0.9pt] (0,0,0) -- (1.6,0,0) node[anchor=north east] {$\langle X \rangle$};
  \draw[->, line width=0.9pt] (0,0,0) -- (0,1.2,0) node[anchor=west] {$\langle Y \rangle$};
  \draw[->, line width=0.9pt] (0,0,0) -- (0,0,1.2) node[anchor=south] {$\langle Z \rangle$};

  \draw[->, line width=1.1pt, green!70!black] (0,0,0) -- (0,0,1)
      node[anchor=west, xshift=3pt, yshift=2pt] {$\vec{H}$};

  \tdplotdrawarc[
      thick, red, line width=0.9pt, ->, >=Stealth
  ]{(0,0,0)}{0.8}{0}{360}{}

  \filldraw[black] (1,1,0.75) circle (0.02);
  \node[anchor=west, xshift=2pt, yshift=2pt] at (1,1,0.8) {$\rho$};

\end{tikzpicture}

\caption{\textbf{Hamiltonian dynamics on the Bloch sphere.} The Hamiltonian in quantum theory (green arrow) acts as the axis of rotation as states (such as $\rho$ represented by the black dot) evolve in the Bloch sphere.}
\label{fig:bloch_rotation}
\end{figure}
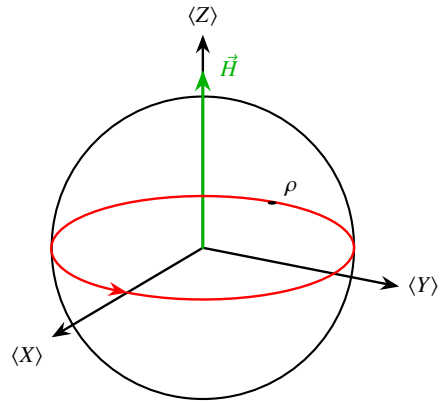

\noindent The corresponding evolution equations for the state vector \( \vec{\rho} = (\rho_1, \rho_2, \rho_3) \) are
\begin{align}
\dot{\rho}_1 &= \rho_2 H_3 - \rho_3 H_2, \label{eq:rho1dot} \\
\dot{\rho}_2 &= -\rho_1 H_3 + \rho_3 H_1, \label{eq:rho2dot} \\
\dot{\rho}_3 &= -\rho_2 H_1 + \rho_1 H_2. \label{eq:rho3dot}
\end{align}
It follows that the quantity \( \vec{H} \cdot \vec{\rho} \) is constant in time, satisfying the desideratum \textbf{INV}. This dynamics is shown in figure \ref{fig:bloch_rotation}. Moreover, this recipe matches the quantum evolution expression derived from $\dot{\rho}=-i\left[\rho,H\right]$\cite{018GenHam}, thereby satisfying \textbf{QUAN}.

This construction can be applied not only to quantum systems but also to non-quantum GPTs with three-dimensional state spaces, such as the gbit mentioned in Section~5.2, the stabilizer polytope ~\cite{012SpeToy}, and Spekkens’ toy model~\cite{007SpeToy}. In each case, the allowable dynamics are restricted by the symmetry group of the state space, but the Hamiltonian can still be defined operationally via the above recipe.

\subsection{Unification of Energy Concepts in Generalised phase space}

In standard quantum theory, the Wigner phase space function plays a crucial role in bridging classical and quantum representations of dynamics by offering a phase-space formulation. The Wigner function is a quasiprobability distribution on phase space that retains many classical-like features, while still encapsulating quantum interference and non-commutativity. GPTs extend this framework beyond both classical and quantum mechanics by investigating arbitrary phase space distributions other than just the Wigner distribution, aiming to capture the essential structure of physical theories more broadly~\cite{plavala2022operational,plavala2022generalized,zachos2005quantum}. Recent work has established how the phase space distributions are associated with energy concepts, offering a unified perspective on dynamical and structural aspects of physical theories~\cite{jiang2024framework,jiang2024unification}.

The dynamics of physical systems in both classical and quantum theories can be expressed within the unified framework of phase space. In classical mechanics, the state of a system is represented by a probability density \(\rho(q, p, t)\) on phase space, and its evolution is governed by the Liouville equation,
\begin{align}
\frac{\partial \rho(q, p, t)}{\partial t} 
&= \{H, \rho\} = -H \Lambda \rho,
\end{align}
where \(H(q, p)\) is the classical Hamiltonian and \(\Lambda = \frac{\overleftarrow{\partial}}{\partial q} \frac{\overrightarrow{\partial}}{\partial p} - \frac{\overleftarrow{\partial}}{\partial p} \frac{\overrightarrow{\partial}}{\partial q}\) is the symplectic operator encoding the structure of Hamiltonian flow. This evolution preserves the normalisation of \(\rho\) and reflects the incompressibility of phase space trajectories under the Hamiltonian dynamics.

In quantum theory, the Wigner function \(W_\rho(q, p, t)\) serves as a quasiprobability representation of the density operator \(\hat{\rho}\). Despite potentially assuming negative values, it retains many classical-like properties and allows for a formulation of quantum dynamics directly in phase space. The time evolution of the Wigner function is governed by the following equation, which can be written in terms of the Moyal star product \(\star = \exp\left( -\frac{i \hbar}{2} \Lambda \right)\) as
\begin{align}
\frac{\partial W_\rho}{\partial t} &= \frac{1}{i\hbar} \left( H \star W_\rho - W_\rho \star H \right) \notag \\
&= -\frac{2}{\hbar} H(q, p) \sin\left( \frac{\hbar}{2} \Lambda \right) W_\rho(q, p),
\end{align}
where \(H(q, p)\) is the Weyl-ordered Hamiltonian. In the classical limit \(\hbar \to 0\), this equation reduces to the Liouville equation, thereby recovering classical dynamics from the quantum formalism.

GPTs extend this picture by allowing more general forms of state distributions and dynamical evolution. In this setting, the dynamics of a state function \(f(q, p, t)\) is described by an equation that unifies the quantum and classical cases:
\begin{equation}
\frac{\partial f}{\partial t} = \int \mathrm{d}k\, K(k)\, f(q, p, t) \sin\left(\frac{k}{2} \Lambda\right) H(q, p),
\end{equation}
where \(K(k)\) is a theory-dependent kernel that characterizes the nature of the dynamical flow. For instance, choosing \(K(k) = \frac{2}{k} \delta(k - \hbar)\) reproduces the quantum equation of motion, while \( \hbar \to 0\) yields the classical Liouville evolution. More generally, arbitrary choices of \(K(k)\) lead to dynamical equations that define postquantum mechanical theories with potentially novel structural features.

\subsection{Negativity in Wigner Distribution}

\begin{figure}[htbp!]
    \centering
    \includegraphics[width=1.0\textwidth]{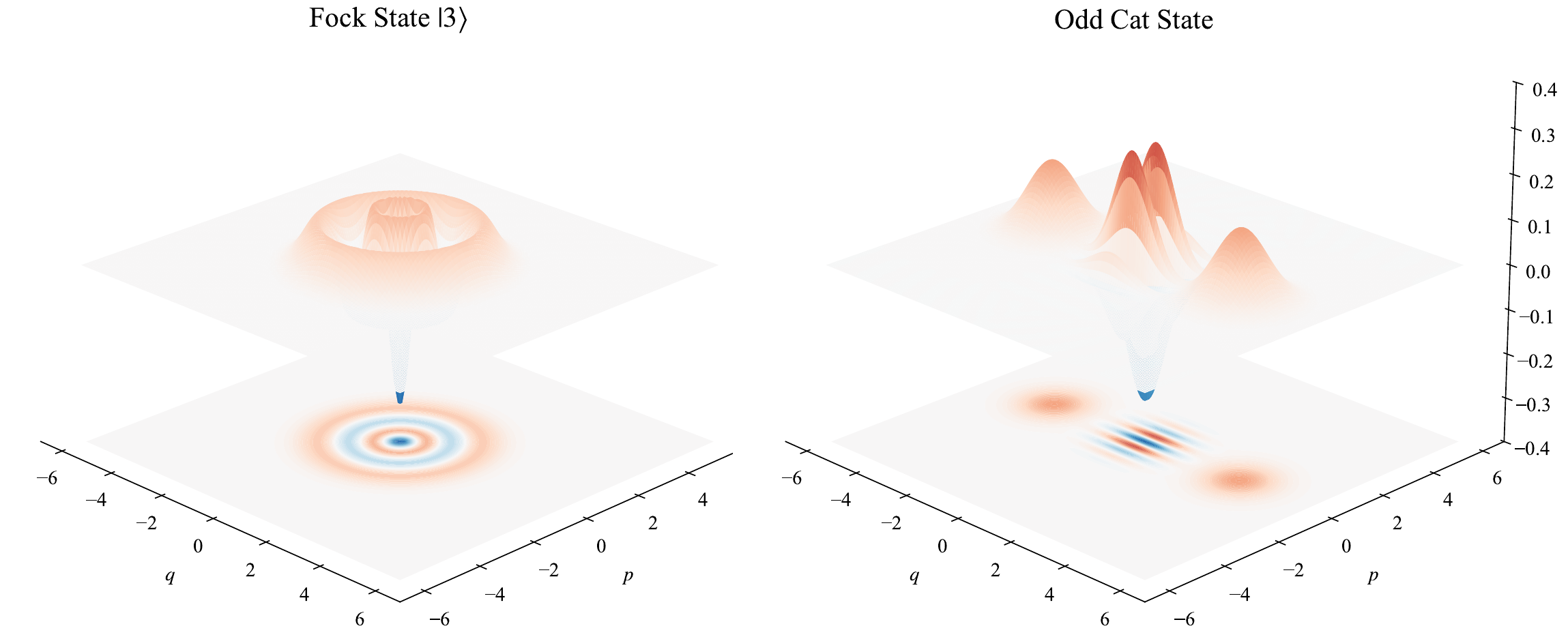}

    \caption{\textbf{Visualization of Wigner function negativity in nonclassical states.} 
    The plots display the Wigner quasi-probability distribution $W(q,p)$ for distinct quantum states. 
    Left: the single-photon Fock state $\ket{3}$. Right: an odd Schrödinger cat state. 
    The blue regions correspond to negative values of the Wigner function ($W < 0$), serving as a direct indicator of nonclassicality that distinguishes these states from classical statistical distributions. 
    The red regions indicate positive probability density.}
    \label{fig:negative_wigner}
\end{figure}

Among various operational phenomena that distinguish quantum mechanics from classical statistical theories, violations of classical probability bounds and quantum tunneling stand out as striking signatures of nonclassicality. Both are directly tied to a characteristic feature of quantum theory: the negativity of the Wigner quasi-probability distribution. To visualize this explicit departure from classical phase-space distributions, Figure~\ref{fig:negative_wigner} illustrates two archetypal quantum states that exhibit clear negative regions. In this subsection, we first discuss violations of classical probability bounds, then examine the relationship between Wigner negativity and contextuality, and finally consider quantum tunneling as another manifestation of negativity. 

\subsubsection{Exceeding classical bounds via Wigner negativity}

A striking signature of nonclassicality in quantum systems is the ability to violate operational bounds that are inviolable under classical probabilistic dynamics. One such example, which we introduce here in detail, concerns the question: How often is the coordinate of a harmonic oscillator positive~\cite{tsirelson2006often}?

In the classical setting, the position of a harmonic oscillator with unit mass and frequency evolves as
\begin{equation}
    q(t) = q_0 \cos t + p_0 \sin t,
\end{equation}
for some fixed initial condition \((q_0, p_0)\). Suppose a measurement of position is performed at a random time \(\tau \in \{0, 2\pi/3, 4\pi/3\} \), with each time equally likely. The coordinate is declared ``positive'' if \( q(\tau) > 0 \). For any fixed phase-space point \((q_0, p_0)\), one can compute \( q(0), q(2\pi/3), q(4\pi/3) \), and check their signs.

A direct analysis confirms that for any classical initial condition \((q_0, p_0)\), the coordinate of a harmonic oscillator can be positive at either one or two of the three specified times \( t = 0, 2\pi/3, 4\pi/3 \), but never all three. Explicitly, the position values at these times are:
\begin{align}
    q(0) &= q_0, \\
    q\left(\frac{2\pi}{3}\right) &= -\frac{1}{2} q_0 + \frac{\sqrt{3}}{2} p_0, \\
    q\left(\frac{4\pi}{3}\right) &= -\frac{1}{2} q_0 - \frac{\sqrt{3}}{2} p_0.
\end{align}
These expressions show that while any one or two of them can be positive depending on the sign and magnitude of \((q_0, p_0)\), it is algebraically impossible for all three to be positive simultaneously. Thus, if the measurement time is chosen uniformly at random from these three, the classical probability of finding the coordinate positive must be either
\begin{equation}
    \text{prob}_{\text{classical}} \in \left\{ \frac{1}{3}, \frac{2}{3} \right\},
\end{equation}
and no other value.

The quantum case is markedly different. In quantum mechanics, the state of the oscillator is described by a wavefunction \(|\psi\rangle\), and the position operator evolves as:
\begin{equation}
    Q(t) = Q \cos t + P \sin t, \quad [Q, P] = i.
\end{equation}
Let
\begin{equation}
    E(t) = \theta(Q(t))
\end{equation}
be the projector onto the positive half-axis of the operator \( Q(t) \). Then the probability of obtaining a positive coordinate is:
\begin{equation}
    \text{prob}(\psi) = \frac{1}{3} \langle \psi | E(0) + E(2\pi/3) + E(4\pi/3) | \psi \rangle.
\end{equation}
Despite the fact that only one measurement is made, the non-commutativity of \( Q(t) \) at different times makes this expression nontrivial. To evaluate it, Tsirelson reformulates the problem using the Wigner quasi-probability distribution.

The Wigner function \( W_\psi(q, p) \) encodes a quantum state as a real-valued function on phase space and satisfies the identity~\cite{tsirelson2006often, bengtsson2017geometry}:
\begin{equation}
    \langle \psi | f(aQ + bP) | \psi \rangle = \iint f(aq + bp)\, W_\psi(q, p)\, dq\, dp.
\end{equation}
In particular, the expectation \( \langle \psi | E(t) | \psi \rangle \) becomes an integral over the half-plane \( q \cos t + p \sin t > 0 \). To simplify the structure further, one defines the \textit{angular Wigner function}:
\begin{equation}
    W_{\text{ang}}^\psi(\varphi) = \int_0^\infty W_\psi(r \cos \varphi, r \sin \varphi)\, r\, dr,
\end{equation}
which represents the radial density of the Wigner function along angle \(\varphi\). The probability then takes the compact form:
\begin{equation}
    \text{prob}(\psi) = \frac{1}{3} \int_{-\pi}^{\pi} f(\varphi)\, W_{\text{ang}}^\psi(\varphi)\, d\varphi,
\end{equation}
where \( f(\varphi) \) is a piecewise constant function taking values 1 or 2 depending on angular sectors.

In a \textit{classical theory}, where the Wigner function is always non-negative, the above integral is necessarily bounded between \( 1/3 \) and \( 2/3 \), matching the earlier conclusion. However, in \textit{quantum mechanics}, the Wigner function may take on negative values. These negative contributions allow the integral to exceed the classical bound. Indeed, a numerical study of the spectrum of the operator \( E(0) + E(2\pi/3) + E(4\pi/3) \) reveals that the maximal achievable probability is approximately:
\begin{equation}
    \text{prob}_{\text{quantum}}^{\text{max}} \approx 0.705 > \frac{2}{3}.
\end{equation}

\noindent In detail, the violation originates from \textit{interference enabled by Wigner negativity}:
\begin{itemize}
    \item The weighting function \( f(\varphi) \) takes value 2 in angular sectors where \textit{two} measurement directions coincide (60° arcs), and 1 elsewhere (120° arcs).
    \item Negative \( W_{\text{ang}}^\psi(\varphi) \) \textit{subtracts} probability in regions where \( f(\varphi) = 1 \).
    \item This ``probability deficit" is \textit{redistributed} to regions where \( f(\varphi) = 2 \).
    \item The net effect is \textit{constructive interference} that boosts the integral beyond \( 2/3 \).
\end{itemize}

\begin{table}
\begin{center}
\begin{tabular}{l|c|c}
 & \textbf{Classical} & \textbf{Quantum} \\ \hline
Phase-space distribution & $\geq 0$ & Can be negative \\  \hline
Interference effects & Absent & Via negativity \\  \hline
\text{prob} $\leq 2/3$ & Always & Violated \\  \hline
Maximal probability & 0.666 & 0.705 \\
\end{tabular}
\end{center}
\caption{Comparison of classical and quantum mechanisms governing the positivity probability, highlighting how Wigner negativity enables violations of the classical upper bound.}
\label{tab:quantum_classical} 
\end{table}


\subsubsection{Negativity and contextuality are equivalent}

Two of the sharpest operational criteria proposed to distinguish quantum theory from classical probabilistic models are \emph{negativity} in quasiprobability representations and \emph{contextuality} in hidden-variable models. Although these were traditionally treated as distinct notions, recent work suggests that the two are closely related, establishing an equivalence when evaluated under operational definitions~\cite{005SpeCon, 008NegCon}.

Historically, the concept of contextuality was formalized through the Kochen--Specker theorem, which examines the possibility of assigning pre-determined values to projective measurements. In this traditional framework, a model is considered \emph{noncontextual} if the value $v(A)$ assigned to an observable $A$ is independent of the context in which it is measured—where a context is defined by a set of mutually commuting observables $\{B, C, \dots\}$ measured alongside $A$. For example, if $A$ commutes with both $B$ and $C$, but $B$ and $C$ do not commute ($[B,C] \neq 0$), a noncontextual model demands that the value $v(A)$ remains identical whether $A$ is measured in the context $\{A, B\}$ or the context $\{A, C\}$.

While foundational, this traditional definition is restricted to sharp (projective) measurements and does not extend to state preparations or, unsharp measurements like POVMs. This limitation motivated the broader operational definition of contextuality proposed by Spekkens. Spekkens generalised this notion by shifting the focus from the assignment of definite outcomes for sharp measurements to the representation of preparations and measurements, and therefore, revealing that contextuality is fundamentally an obstacle to representing quantum theory with a positive-valued probability measure.

In a quasiprobability representation, every quantum state $\rho$ is associated with a real, normalised function
\begin{equation}
\rho \;\longleftrightarrow\; \mu_\rho(\lambda), 
\quad \int d\lambda\, \mu_\rho(\lambda) = 1,
\end{equation}
and $\lambda$ is a variable that determines the distribution. For every POVM element $E_k$, it is associated with a real function
\begin{equation}
E_k \;\longleftrightarrow\; \xi_{E_k}(\lambda),
\quad \sum_k \xi_{E_k}(\lambda)=1 \;\;\forall \lambda.
\end{equation}
These assignments must reproduce the Born rule:
\begin{equation}
\mathrm{Tr}(\rho E_k) = \int d\lambda\, \mu_\rho(\lambda)\,\xi_{E_k}(\lambda).
\label{Born-qp}
\end{equation}
A representation is called \emph{nonnegative} if
\begin{equation}
\mu_\rho(\lambda) \geq 0, \qquad 
\xi_{E_k}(\lambda) \geq 0,
\label{positivity-qp}
\end{equation}
for all $\rho$ and all POVM elements $E_k \leq I$. If such a representation exists, the theory admits a purely classical interpretation; the failure of~\eqref{positivity-qp} (i.e., the appearance of negative values) is then a hallmark of nonclassicality.

To provide a realist explanation for these equations, one appeals to an ontological model. In this framework, the physical system is assumed to possess objective attributes described by an ontic state $\lambda$ in a space $\Omega$, which is independent of observations. If the preparation procedure does not uniquely determine $\lambda$, the additional degrees of freedom are known as hidden variables.

In the ontological picture, a \emph{preparation procedure} $P$ is represented by a probability distribution over these states:
\begin{equation}
P \;\longleftrightarrow\; \mu_P(\lambda), 
\quad \mu_P(\lambda) \geq 0,\;\; 
\int d\lambda\, \mu_P(\lambda)=1,
\end{equation}
while a \emph{measurement procedure} $M$ is represented by a set of response functions—the probability of obtaining outcome $k$ given the system is in state $\lambda$:
\begin{equation}
M \;\longleftrightarrow\; \{\xi_{M,k}(\lambda)\}, 
\quad \xi_{M,k}(\lambda)\geq 0,\;\;
\sum_k \xi_{M,k}(\lambda)=1.
\end{equation}
To be empirically viable, the model must reproduce the observed statistics of quantum theory via the Born rule:
\begin{equation}
\int d\lambda\, \mu_P(\lambda)\,\xi_{M,k}(\lambda) 
= \mathrm{Tr}(\rho E_k) = \langle \psi| E_k |\psi \rangle,
\label{Born-ont}
\end{equation}
for every $P$ implementing the state $\rho$ and every $M$ implementing the POVM $\{E_k\}$. This equation acts as the bridge between the hidden gears of the ontological model and the operational results we see in the lab.

In the generalised framework proposed by Spekkens~\cite{005SpeCon, 008NegCon}, the assumption of noncontextuality can be summarized by the principle that ``identical behavior in the lab implies identical reality in the model". In other words, any operational equivalence at the macroscopic level must be reflected as an identity at the ontological level. 

To formalize this, consider two preparation procedures, $P$ and $P'$, that yield identical statistics for all possible measurements. They are operationally equivalent ($P \simeq P'$) and can be associated with the same state $\rho$. The specific choice of procedure $P \in \mathcal{P}_\rho$ constitutes the preparation context. Noncontextuality demands that the model cannot differentiate between these contexts; it must assign the same ontic state to any procedure in the equivalence class.

Similarly, let $M$ be a full measurement apparatus (e.g., a specific choice of POVM) containing a measurement outcome $k$. If an outcome in measurement $M$ yields the exact same statistics as an outcome in a different measurement $M'$ for all prepared states, they are operationally equivalent and correspond to the same quantum effect $E_k$. Here, the choice of the full apparatus $M \in \mathcal{M}_{\{E_k\}}$ constitutes the measurement context.

Mathematically, a universally noncontextual ontological model requires that the hidden distributions depend only on these operational equivalence classes ($\rho$ and $E_k$), rather than the specific experimental contexts ($P$ and $M$):
\begin{align}
\mu_P(\lambda) &= \mu_\rho(\lambda) 
& \text{for all } P\in \mathcal{P}_\rho, \label{prep-nc}\\
\xi_{M,k}(\lambda) &= \xi_{E_k}(\lambda) 
& \text{for all } M\in \mathcal{M}_{\{E_k\}}. \label{meas-nc}
\end{align}

This mathematical formulation highlights exactly how the traditional Kochen--Specker notion emerges as a special case of Spekkens' framework~\cite{005SpeCon,008NegCon}. While the original Kochen--Specker theorem relies strictly on sharp, projective measurements, Spekkens' operational framework applies to arbitrary procedures, including POVMs. When one restricts focus to projective measurements and further imposes the classical assumption of outcome determinism, the response functions for a projector $\Pi_k$ must satisfy $\xi_{\Pi_k}(\lambda) \in \{0,1\}$. Here, $M$ and $M'$ denote distinct projective measurement contexts containing the same projector $\Pi_k$, corresponding to different sets of mutually commuting observables measured jointly with $\Pi_k$. In this deterministic limit, Spekkens' measurement noncontextuality condition, $\xi_{M,k}(\lambda) = \xi_{M',k}(\lambda)$, reduces to the Kochen--Specker requirement that the predetermined value assigned to $\Pi_k$ is independent of the commuting observables that complete the measurement context.

To see the direct connection to quasiprobability representations, we can substitute these noncontextuality conditions back into the ontological Born rule from Eq.~\eqref{Born-ont}. This yields:
\begin{equation}\int d\lambda \mu_\rho(\lambda)\xi_{E_k}(\lambda) = \mathrm{Tr}(\rho E_k).
\label{Born-bridge}
\end{equation}
Notice what has happened here: the explicit experimental context labels $P$ and $M$ have vanished. The functions now depend strictly on the operational quantum state $\rho$ and the measurement operator $E_k$. Because an ontological model inherently requires its functions to be non-negative ($\mu \geq 0$ and $\xi \geq 0$), Eq.~\eqref{Born-bridge} is mathematically identical to the definition of a non-negative quasiprobability representation given in Eq.~\eqref{Born-qp} and Eq.~\eqref{positivity-qp}.

This structural identity in Spekkens' work equates a noncontextual ontological model with a non-negative quasiprobability representation. Because quantum theory cannot be modeled within a strictly non-negative, noncontextual framework, any attempt to replicate its predictions faces an inevitable mathematical trade-off. Preserving a noncontextual representation requires dropping the explicit context labels $P$ and $M$, which forces the phase-space functions $\mu_\rho(\lambda)$ or $\xi_{E_k}(\lambda)$ to take on negative values. Conversely, enforcing strict non-negativity requires the model to retain these experimental labels, yielding a contextual representation where operationally identical procedures map to distinct ontic distributions ($\mu_P(\lambda) \neq \mu_{P'}(\lambda)$). Consequently, quasiprobability negativity and operational contextuality are not distinct physical phenomena, but rather mathematically equivalent expressions of the same underlying non-classical properties.

\subsubsection{Wigner functions and tunneling in GPT}

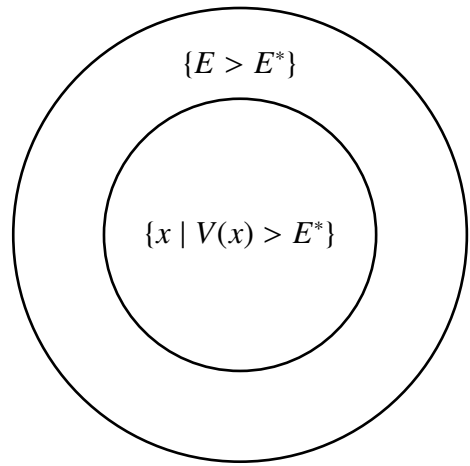
\begin{figure}[htbp!]
    \centering
    \begin{tikzpicture}[scale=2.0, line cap=round, line join=round, >=Stealth, font=\normalsize]

        \def\rout{1.5}
        \def\rin{0.9}

        \draw[line width=1pt] (0,0) circle (\rout);
        \draw[line width=1pt] (0,0) circle (\rin);

        \node at (0, 0.75*\rout) {\Large $\{E > E^*\}$};
        \node at (0, 0) {\Large $\{x \mid V(x) > E^*\}$};

    \end{tikzpicture}

    \caption{\textbf{Classically forbidden regions and quantum tunneling.} In classical mechanics, regions with $V(x) > E^*$ are inaccessible unless the particle has energy $E > E^*$. In quantum mechanics, however, the particle with energy $E \leq E^*$ could enter this region. The Wigner function is used to explain this, supported in part by negative quasiprobabilities that enable tunneling.}
    \label{fig:tunneling_negative_quasiprobability}
\end{figure}

Quantum tunneling is one of the most iconic examples of behavior that has no known classical analogue. Classically, a particle with total energy \(E\) cannot be found in any region where the potential energy \(V(x)\) exceeds \(E\). Yet quantum particles do exactly this: they may be found in classically forbidden regions with nonzero probability, even when prepared in energy eigenstates with \(E < V(x)\). To generalise this concept beyond quantum theory, we adopt an operational definition of tunneling that is compatible with the Wigner function formalism and extendable to generalised probabilistic theories~\cite{lin2020necessity, razavy2013quantum, marinov1996quantum, kriman1987scattering}.



A more general and operational definition is the following: a state exhibits tunneling if there exists some threshold energy \(E^*\) such that the probability of finding the particle in a region where the potential energy exceeds \(E^*\) is greater than the probability that the measured energy of the state exceeds \(E^*\). That is, tunneling occurs whenever~\cite{lin2020necessity}
\begin{equation}
    P(x \in \{ x : V(x) > E^* \}) > P(E > E^*).
\end{equation}
The scenario is also shown by figure \ref{fig:tunneling_negative_quasiprobability}. This definition does not require the state to be an energy eigenstate and applies to arbitrary states and potentials.

To analyze this in phase space, we turn to the Wigner function formulation. In this picture, a quantum state is represented by a real-valued function \(W(x, p)\) on phase space, and measurement probabilities take the form of integrals over \(x\) and \(p\). Specifically, the probability of a measurement outcome associated with an effect operator is given by the overlap integral:
\begin{equation}
    P = \iint W(x, p) \cdot W_{\text{effect}}(x, p) \, dx \, dp,
\end{equation}
where \(W_{\text{effect}}(x, p)\) is the Wigner representation of the measurement effect. This structure closely resembles classical mechanics, where probabilities are computed via phase space integrals using delta functions. For example, in classical mechanics, the probability of a quantity \(\mathcal{O}(x, p)\) equaling some value \(\omega\) is written as:
\begin{equation}
    P = \iint f(x, p) \cdot \delta[\mathcal{O}(x, p) - \omega] \, dx \, dp,
\end{equation}
where \(f(x, p)\) is the classical probability density. In this sense, delta functions in classical theory play a role analogous to the Wigner functions of measurement operators in quantum theory.

To make the general structure of tunneling explicit, consider a system whose state is represented by a phase-space distribution \(f(x, p)\). Let \(\varepsilon_{E > E^*}(x, p)\) denote the effect function corresponding to the probability of measuring total energy greater than a threshold \(E^*\), and let \(\varepsilon_{x\,|\,V(x) > E^*}(x, p)\) denote the effect function corresponding to finding the particle in a region where the potential \(V(x)\) exceeds \(E^*\). The probability of each event is obtained by integrating the distribution \(f(x, p)\) against the respective effect:
\begin{equation}
    P(E > E^*) = \iint \varepsilon_{E > E^*}(x, p)\, f(x, p)\, dx\, dp,
\end{equation}
\begin{equation}
    P(x \in \{x : V(x) > E^*\}) = \iint \varepsilon_{x\,|\,V(x) > E^*}(x, p)\, f(x, p)\, dx\, dp.
\end{equation}

A state is said to exhibit tunneling if there exists some threshold \(E^*\) for which the probability of the system being located in a region where \(V(x) > E^*\) exceeds the probability that its total energy is greater than \(E^*\). In other words, tunneling occurs if and only if
\begin{equation}
    \iint \left[ \varepsilon_{E > E^*}(x, p) - \varepsilon_{x\,|\,V(x) > E^*}(x, p) \right] f(x, p)\, dx\, dp < 0.
\end{equation}

This inequality defines tunneling as a property of the joint structure of the state and the measurement effects in phase space. In classical probability theory, where \(f(x, p)\) is a proper distribution and both \(\varepsilon\) functions are non-negative, this integral is always non-negative. Hence, a violation of this inequality is a signature of nonclassicality.

In the quantum case, we identify \(f(x, p)\) with the Wigner function \(W(x, p)\), and interpret \(\varepsilon_{E > E^*}(x, p)\) and \(\varepsilon_{x\,|\,V(x) > E^*}(x, p)\) as the Wigner representations of the corresponding measurement effects. The integral then becomes:
\begin{equation}
    \iint \left[ W_{\varepsilon_{E > E^*}}(x, p) - W_{\varepsilon_{x\,|\,V(x) > E^*}}(x, p) \right] W(x, p)\, dx\, dp < 0.
\end{equation}

This shows that tunneling occurs if and only if the overlap between the state and the difference of these two effect functions is negative. Because all terms now lie within the Wigner formalism, any violation of this inequality must originate from the presence of \textit{negativity} in at least one of the functions involved.

To illustrate this result, consider the quantum harmonic oscillator. The ground state is a pure Gaussian state whose Wigner function is strictly positive, given by
\begin{equation}
    W_0(x, p) = \frac{1}{\pi} \exp\left( -x^2 - p^2 \right).
\end{equation}
Nonetheless, under the operational definition of tunneling described above, the ground state still qualifies as a tunneling state. The reason is that while the state's Wigner function is entirely positive, the Wigner representation of the energy measurement effect—specifically, the operator projecting onto energies greater than the ground state energy \(E^* = \frac{1}{2}\)—is negative in certain regions of phase space. Their overlap is given by
\begin{equation}
    \iint \left[ W_{\varepsilon_{E > \frac{1}{2}}}(x, p) - W_{\varepsilon_{x\,|\,V(x) > \frac{1}{2}}}(x, p) \right] W_0(x, p)\, dx\, dp < 0,
\end{equation}
which confirms the presence of tunneling. This demonstrates that even seemingly classical states can tunnel—not because the state itself is nonclassical in Wigner terms, but because the measurement is.

This example highlights the nuanced and foundational role played by Wigner function negativity. Tunneling is not just an outcome of wave mechanics or potential barrier shapes—it is a manifestation of nonclassical probability structure. In the broader framework of GPTs, such behavior distinguishes quantum theory from any theory with fully classical phase-space structure. It provides a universal signature of nonclassicality that is independent of specific Hamiltonians.

\subsection{Further readings}



GPTs offer a natural extension of quantum and classical mechanics, encompassing reversible dynamics and providing a generalised view on entropy and randomisation~\cite{Hardy,Barrett07,Mielnik74}. Recent works on entropy and information in GPTs~\cite{mueller2012black,muller2012unifying,short2010entropygpt,barnum2010entropycausality} suggest these frameworks can be applied to the black hole information problem. For instance, within GPTs, the release of information during black hole evaporation can be described in a way that differs from standard quantum theory, allowing for the possibility of rapid information emission without violating fundamental principles like the no-hiding theorem~\cite{braunsteinciphers,how-merge2,fqsw,braunstein-pati-nohiding,uhlmann1976transition}. These generalisations could offer an alternative to black hole complementarity, helping to resolve the tension between information preservation and thermal radiation.

To further explore the framework presented in this section, we briefly mention some key papers related to Hamiltonians in GPTs, negativity in Wigner distributions, and contextuality. For Hamiltonian mechanics in GPTs, Barnum \textit{et al.}~\cite{barnum2014higher} explore higher-order interference and its role in characterising quantum theory. Generalising phases like those in quantum time evolution and interferometers has been discussed by Garner et al.\cite{013GPTPha, garner2018interferometric}.  In the context of negativity in Wigner distributions, Marinov and Segev~\cite{Marinov1996} discuss quantum tunneling, while Kriman \textit{et al.}~\cite{Kriman1987} and Roy and Khan~\cite{Roy1993} focus on scattering states and tunneling in microstructures. On contextuality, Lütkenhaus and Barnett~\cite{Lütkenhaus1995} study nonclassical effects in phase space, Bell~\cite{Bell1966} addresses hidden variables in quantum mechanics, and Spekkens~\cite{Spekkens2005} analyses contextuality in preparations and unsharp measurements. These papers offer valuable perspectives and deeper insights into the discussed topics, and readers may find them useful for further exploration.

\bibliographystyle{JHEP}%
\bibliography{refsInTemplate}

\end{document}